\documentclass[manuscript]{aastex}
\usepackage{graphicx} 
\usepackage{amsmath}
\usepackage{xcolor}

\shorttitle{Model predictions for future heliospheric ENA imagers}
\shortauthors{Galli et al.}

\begin{document}

\title{An empirical model of Energetic Neutral Atom imaging of the heliosphere and its implications
for future heliospheric missions at great heliocentric distances}

\author{A. Galli and P. Wurz}
\affil{Physics Institute, University of Bern,
    Bern, Switzerland}

\author{H. Fichtner}
\affil{Ruhr University, Bochum, Germany}

\author{Y. Futaana and S. Barabash}
\affil{IRF Swedish Institute of Space Physics, Kiruna, Sweden}


\begin{abstract} 
Several concepts for heliospheric missions operating at heliocentric distances far beyond Earth orbit
are currently investigated by the scientific community.
The mission concept of the Interstellar Probe \citep{mcn18}, e.g., aims at reaching a distance of 1000 au away 
from the Sun within this century. This
would allow the coming generation to obtain a global view of our heliosphere from an outside
vantage point by measuring the 
Energetic Neutral Atoms (ENAs) originating from the various plasma regions. It would also allow for direct sampling
of unperturbed interstellar medium, and for many observation opportunities beyond heliospheric science, such
as visits to Kuiper Belt Objects, a comprehensive view on the interplanetary dust populations, and 
infrared astronomy free from the foreground emission of the Zodiacal cloud.

In this study, we present a simple empirical model of ENAs from the heliosphere and 
derive basic requirements for ENA instrumentation onboard a spacecraft at great heliocentric distances.
We consider the full energy range of heliospheric ENAs from 10 eV to 100 keV because
each part of the energy spectrum has its own merits for heliospheric science. 
To cover the full ENA energy range, two or three different ENA instruments are needed.    
Thanks to parallax observations, some insights about the nature of the IBEX 
Ribbon and the dimensions of the heliosphere can already be gained by ENA 
imaging from a few au heliocentric distance. To directly reveal the global shape
of the heliosphere, measurements from outside the heliosphere are, of course, the best option.

\end{abstract}


\section{Introduction}\label{sec:introduction}

In this study, we investigate basic requirements for Energetic Neutral Atom (ENA) instruments on a spacecraft
headed for heliocentric distances beyond Mars orbit for the ENA energy range between 10 eV and 100 keV. 
An ENA is produced when an ion of a plasma population exchanges its 
charge with an ambient neutral atom \citep{roe85,gru01}. The resulting ENA leaves its source region on a ballistic trajectory, 
no longer influenced by electromagnetic fields. This allows an ENA camera to image the ion distribution of remote plasma regions \citep{wur00,fah07}.

\begin{figure}
  \includegraphics[width=1.0\textwidth]{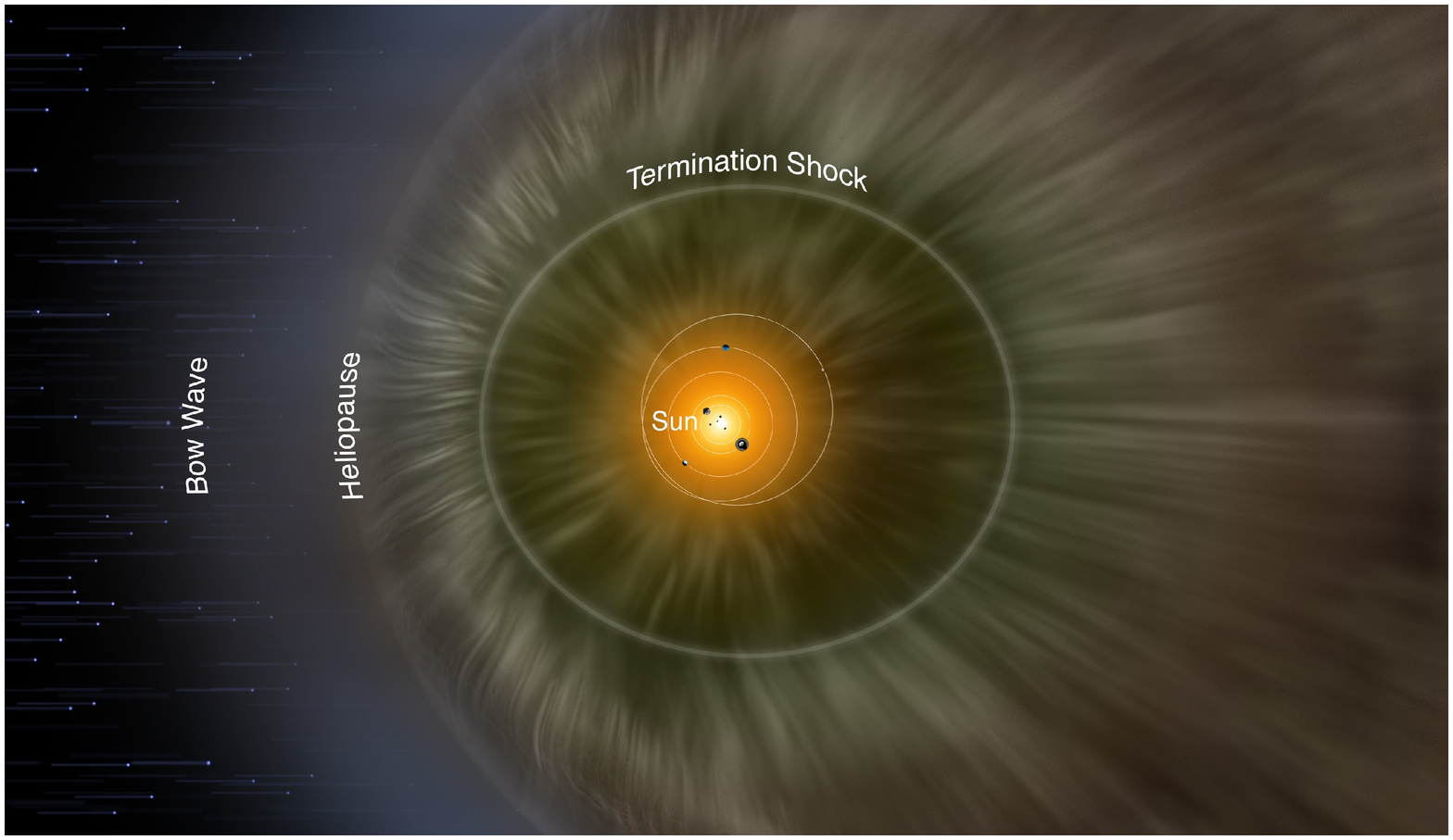}
\caption{Artist's impression of the heliosphere. Image Credit: IBEX Team/Adler Planetarium}\label{fig:artistic}
\end{figure}

Because of its remote sensing character, ENA imaging is an indispensable method to derive a global
view of the heliosphere. The heliosphere is the vast plasma region of the solar wind expanding against 
the surrounding interstellar medium (ISM). The ISM flows past 
the heliosphere at roughly 25 km s$^{-1}$ \citep{mcc15}. This relative flow speed defines
an upwind and a downwind direction in the heliosphere. The outer plasma boundary
between the solar wind and the ISM is called the heliopause. 
The two Voyager spacecraft have crossed this boundary in upwind direction 
at roughly 120 au from the Sun \citep{sto13}. The heliosphere itself is divided by the termination
shock \citep{bur05,bur08} into an inner region of supersonic solar wind and an outer region
of shocked solar wind (the inner heliosheath). Our current knowledge or lack thereof of the heliosphere 
is illustrated by the artist's impression in Fig.~\ref{fig:artistic}: 
the shape and dimension of the heliosphere in the downwind hemisphere are unknown (see Section~\ref{sec:shapes}),
and the existence of a bow wave or a bow shock in the ISM approaching the 
heliopause is still a matter of debate \citep{mcc12,zan13,sche14}. 
Our knowledge about the heliosphere owes a lot to in-situ plasma measurements and remote ENA imaging performed at 
1 au (see Section~\ref{sec:proton_distributions}),
but some questions about the global heliosphere shape, plasma populations and pressure balances 
beyond the termination shock may be impossible to answer with observations restricted to the inner solar system.

A NASA-funded study currently investigates the scientific and technical requirements for an 
Interstellar Probe with the goal of reaching 1000 AU within 50 years. 
The science targets include exploration of the heliosphere and its interaction with the ISM, 
characterization of the circum-solar dust disk, exploration of Kuiper Belt Objects, and astronomical observations in the infrared
wavelength range beyond the zodiacal cloud \citep{mcn18,bra19}.
On a similar note, the China National Space Administration is investigating a mission scenario in 
which two ``Heliospheric Boundary Explorers'' would be launched towards the upwind and towards the downwind direction
of the heliosphere, respectively, to reach 100 au distance by 2049 \citep{zon18}. For ESA's Cosmic Vision, the Local Interstellar Medium Observatory
(LIMO) was proposed to accurately sample the interstellar neutral gas and dust 
and to measure heliospheric ENAs at $1^{\circ}$ angular resolution at a heliocentric distance of 3 au, thus avoiding 
many complications introduced by solar gravity and radiation closer to the Sun \citep{bar19}.

A spacecraft at great heliocentric distance obviously allows for many new ground-breaking 
measurements in addition to ENA imaging. In this study, we limit ourselves
to the prediction and discussion of heliospheric ENAs:
we have integrated the existing measurements of heliospheric ENAs from the heliosheath and beyond
into a simple empirical model. This allows us to make some predictions for an ENA instrument onboard
a spacecraft at heliocentric distances beyond Mars orbit.  
Interstellar neutral (ISN) He, H, and other ISN species flowing into the heliosphere 
\citep{mue04,wit04,moe12,rod12,sau12,kub14,mcc15,par16,gal19} could also
be measured with ENA cameras such as IBEX-Lo \citep{fus09} 
or with a mass spectrometer \citep{bar19}.
However, we have to relegate questions about ISN observation strategies to a future publication
because our heliosphere model currently does not include 
ISN trajectories. The ISN hydrogen is just modeled
as a static density relevant for ENA sources and losses.
Measuring ISN will be an important goal for a heliospheric mission bound to 
heliocentric distances beyond 1 au: the effects of the Sun's gravity, solar wind pressure, and 
ionization rates drop with the square of the distance to the Sun, thus ISN measurements at several au or beyond
will closely resemble the ISN at the heliopause in terms of energy, direction, and composition.
A fast moving spacecraft ($\geq 25$ km s$^{-1}$) must be heading to a vantage point in the upwind hemisphere
to seize this opportunity for ISN sampling.    


\section{The empirical model of heliospheric ENAs}\label{sec:model}

We have created a simple empirical model to predict the differential intensity $j_{\textup{ENA}}$
(in units of cm$^{-2}$ sr$^{-1}$ s$^{-1}$ keV$^{-1}$) of heliospheric
ENAs an ENA camera would observe for any given location and viewing direction.
The fundamental ENA equation,

\vspace{-7mm}

\begin{equation}
j_{\textup{ENA}}(E) = \int_{\textup{LOS}} dl \, (j_p(E)\, n_{\textup{H}}\, \sigma(E)) - L(E),\label{eq:ENA}
\end{equation}

\noindent tells us that a hydrogen ENA model requires the local proton intensity $j_p(E)$, neutral densities $n_{\textup{H}}$,
charge-exchange cross-sections $\sigma(E)$ for the reaction H$^+ +$ H $\rightarrow$ H$^*$ + H$^+$
and the spatially variable ENA loss term $L(E)$ to predict $j_{\textup{ENA}}$ for a given line-of-sight (LOS). 

Equation~\ref{eq:ENA} is a simplification for several reasons:
Most importantly, we only consider hydrogen ENAs here. The second most abundant heliospheric 
species, helium ENAs from neutralized solar wind and pick-up ions, is expected to generate 
an intensity of typically 1 cm$^{-2}$ sr$^{-1}$ s$^{-1}$ keV$^{-1}$ at 1 au for optimum observation
conditions, i.e., 1 keV ENA energy and observing the downwind region \citep{swa17}. 
This He signal is two orders of magnitude weaker than typical H 
ENA intensities at 1 au and has not been observed yet, but it should be detectable with the future IMAP-Hi instrument \citep{swa17,mcc18}. The long cooling length of keV helium ions in the heliosheath would make He ENAs ideally suited to probe the far heliosheath \citep{swa17}. 

We also assume that the hydrogen ENAs are moving along straight lines as soon as they have been created. 
Gravity forces and the influence of solar UV radiation and solar wind are therefore neglected. 
This is acceptable as long as we consider only ENAs with energies of at least 100 eV or heliocentric distances 
greater than 1 au. The sum of these effects would change the observed energy of 100 eV ENAs observed at 1 au by less
than 10\% and the deflection angle relative to the original trajectory would be at most $1^{\circ}$ at 1 au 
for 100 eV ENAs (and even smaller for higher energies) for any solar wind conditions \citep{bzo08}. 
For a 20 eV ISN H atom traveling towards the Sun, deviations from a straight trajectory become notable 
only for heliocentric distances less than 3 au for any solar conditions \citep{rah19}.

Finally, neutral species other than hydrogen could in principle also neutralize protons. 
However, helium as the most abundant non-hydrogen species
has a density of only $n_{\textup{He}}\simeq 0.01\dots0.02$ cm$^{-3}$ everywhere inside and beyond the heliopause \citep{glo04,mue04}.
This is one order of magnitude less than $n_{\textup{H}}$. Because the charge-exchange cross-section for the reaction H$^+ +$ He $\rightarrow$ H$^*$ + He$^+$  is also at least one order of magnitude smaller than for H$^+ +$ H $\rightarrow$ H$^*$ + H$^+$ at all energies below 
10 keV \citep{bar90,lin05}, the neutral helium would only have to be considered for ENAs of energies exceeding 100 keV.

In the following, we will present how we implemented the individual terms in Equation~\ref{eq:ENA} and their underlying assumptions in more detail:

\subsection{Heliospheric shapes}\label{sec:shapes}

First, we must define closed three-dimensional shapes for the termination shock and for the heliopause
to decide which local ion populations to consider for the line-of-sight integration in Eq.~\ref{eq:ENA}.
Whether a bow shock forms around the heliopause or not is still debated \citep{mcc12,zan13,sche14}, 
but is irrelevant for our current model.

Our shape of the termination shock conforms to the observational constraints from Voyager 1 and 2
(they crossed the termination shock toward upwind direction at heliocentric distances of 94 and 83 au \citep{bur05,bur08})
and are consistent with IBEX ENA spectra (see \citet{sch11,rei16,gal16}). 
We model the termination shock as an ellipsoid whose center is shifted with respect to the
Sun:

\vspace{-7mm}

\begin{equation} 
\left( \begin{array}{lll} x \\ 
y \\
z\end{array} \right)  = \left( \begin{array}{lll} a_e \cos(\vartheta) \cos(\varphi) + x_e \\
                              b_e \cos(\vartheta) \sin(\varphi) + y_e\\  \label{eq:ellipsoid}
                              c_e \sin(\vartheta) + z_e \end{array} \right),
\end{equation}

\noindent The dimensional parameters read (in au) 
$a_e = 100$, $b_e = 100$, $c_e = 120$, $x_e = -15$, $y_e = 0$, and $z_e = 0$, 
which implies heliocentric distances towards the poles of 120 au and 85 au (upwind) and 115 au (downwind) in the $xz$-plane. 

\begin{figure}
  \includegraphics[width=1.0\textwidth]{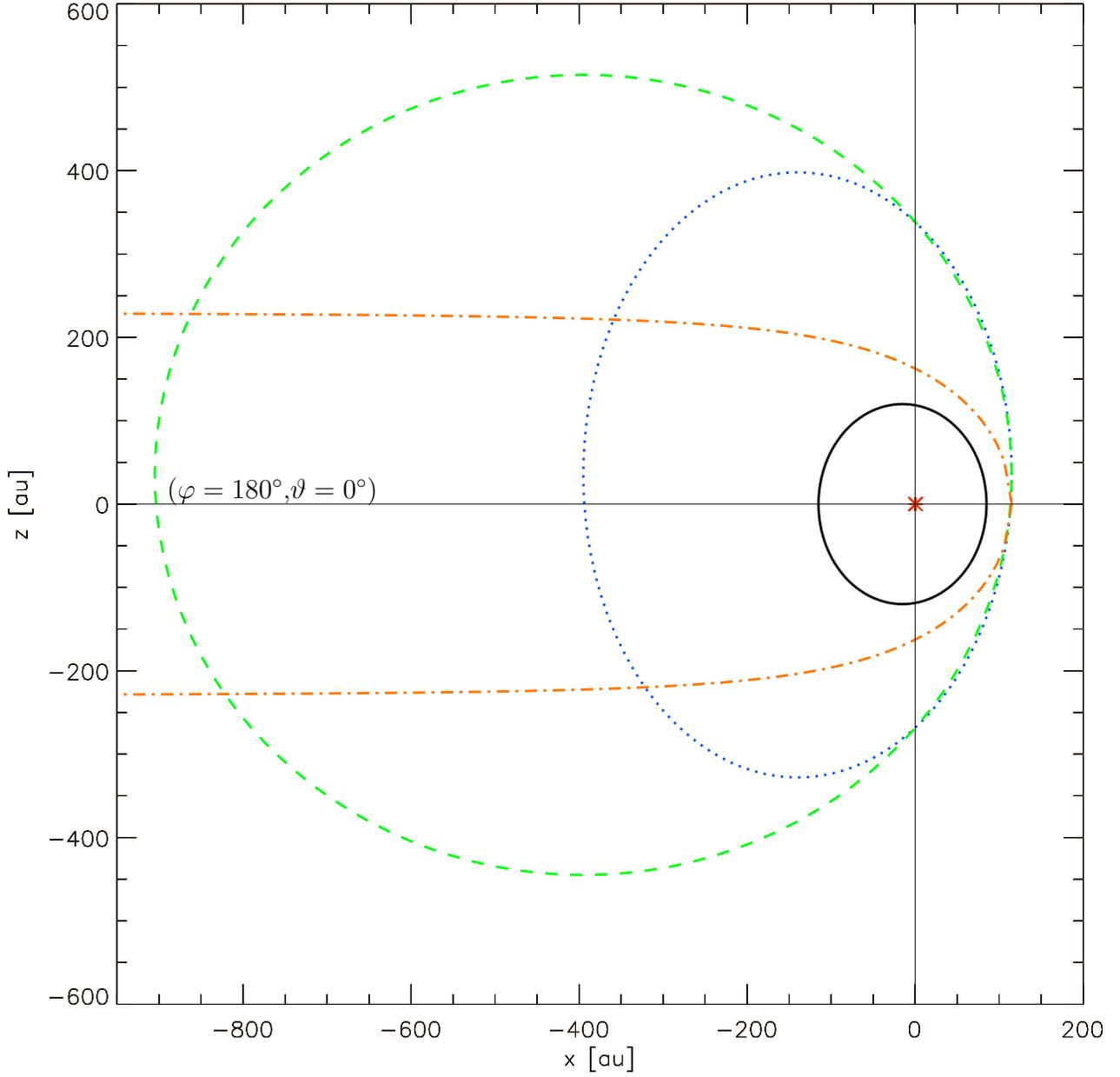}
  {\put(-400,250){($\varphi=180^{\circ}$,$\vartheta=0^{\circ}$)}}
\caption{Shapes of the termination shock (black solid line) and the three different heliopause cases
(blue dotted line: small ellipsoid, green dashed line: large ellipsoid, orange dashed-dotted: cylindrical Parker model) 
assumed in this study. The Sun (red asterisk) is situated at
the center of the coordinate system, the $x$-axis from the Sun to the nose of the heliosphere
points to $\varphi=0^{\circ}$, $\vartheta=0^{\circ}$ in polar coordinates.}\label{fig:shapes}
\end{figure}

Throughout this paper, we use spherical coordinates with longitude $\varphi$, latitude $\vartheta$, and heliocentric 
distance $R$. The reference frame is a rotated ecliptic coordinate system centered in the Sun:
The $x$-axis points from the Sun to the nose or upstream direction ($\varphi = 0^{\circ}$, $\vartheta=0^{\circ}$), 
corresponding to $\lambda \approx 256^{\circ}$, $\beta \approx 5^{\circ}$ in ecliptic longitude and latitude \citep{mcc15}.
The $z$-axis lies inside the plane spanned by the $x$-axis and the North pole of the ecliptic 
(i.e., $5^{\circ}$ offset from the ecliptic North pole), and the $y$-axis closes the right-handed system.    
An illustration of this coordinate system and the heliospheric shapes in the $xz$-plane are shown in Fig.~\ref{fig:shapes}.

The global shape of the heliosphere outside the nose region is unknown; for an overview of the possible solutions see, 
e.g., \citet{pog17} and \citet{oph16a} and also consider \citet{dia17} and \citet{sch18}. We therefore experiment with three different heliopause shapes: 
they all conform with the upwind stand-off distance
known from the Voyager crossings at 122 au and 119 au \citep{sto13}.
Two of the heliopause shapes are ellipsoids with an offset relative to the Sun
in analogy to the termination shock (also see \citet{fah86}). The dimensional parameters
in Equation~\ref{eq:ellipsoid} for our case of a ``small ellipsoid'' heliopause read in units of au: 
$a_e = 255$, $b_e = 358$, $c_e = 363$, $x_e = -140$, $y_e = 0$, and $z_e = 35$.
These parameters become $a_e = 510$, $b_e = 474$, $c_e = 480$, $x_e = -395$, $y_e = 0$, and $z_e = 35$
for the case of a ``large ellipsoid'' with an extended heliosheath region in downwind direction.
In addition to these ellipsoids, we also implemented the cylindrical Parker shape for the heliopause
\citep{par61,roe15}:

\vspace{-7mm}

\begin{equation} 
\left( \begin{array}{lll} x \\ 
y \\
z\end{array} \right)  = \left( \begin{array}{lll} (2 a_e^2-\rho^2)/(\sqrt{4 a_e^2-\rho^2}) \\
                              \rho \cos(\varphi)\\  \label{eq:HP_cylindrical}
                              \rho \sin(\varphi) \end{array} \right),
\end{equation}
\noindent with the standoff distance $a_e = 115$ au at the $x$-axis and the cylindrical
coordinates $\varphi$ and $\rho$ (radial distance from $x$-axis). The cylindrical shape
and the ellipsoids are symmetric with respect to the $x$-axis. They can thus 
be characterized by their cross-sections in the ${x,z}$-plane as illustrated in Fig.~\ref{fig:shapes}.
Additional geometric shapes for the termination shock or the heliopause could easily be implemented, but
we will restrict the discussion in Section \ref{sec:applications} to these three cases as they cover a wide range
of possible geometries. 
The inner heliosheath thickness of $150-220$ au towards the poles for the ellipsoids is motivated by \citet{gal16} and \citet{rei16}.
A re-analysis of the method from \citet{rei16} showed that the temporal variations of ENA intensities imply
a much shorter heliosheath thickness of only 50 au towards the poles \citep{rei19}. 
This value is consistent with the cylindrical shape (see Fig.~\ref{fig:shapes}).

\subsection{Neutral densities} 

Inside the heliopause, we assume a constant neutral hydrogen density $n_H = 0.1$~cm$^{-3}$. 
This is correct within a factor of 2 \citep{sch11,hee14,glo15} and, because we scale the
ion intensities in such a way as to reproduce the observed ENA intensities, the explicit value
is irrelevant in our code. The same holds true for ENA sources in the outer heliosheath. Here, $n_H$ would be 0.2 cm$^{-3}$ 
within a factor of 2 for all regions including the hydrogen wall \citep{mue04,glo11,hee14,oph16}.    

\subsection{Charge-exchange cross-sections} 

The charge-exchange cross-section $\sigma(E)$ depends on energy but is well known \citep{bar90,lin05}. 
We rely on the semi-empirical formula by \citet{lin05} to calculate $\sigma(E)$ for the 
reaction H$^{+}$ + H $\rightarrow$ H$^*$ + H$^{+}$.

\subsection{ENA loss processes}

ENA loss processes are so far included in a crude manner: inside the heliopause,
ENA losses are neglected. More precisely, the actual ENA losses amount to less than 10\%
for the proton densities in the inner heliosheath of $n_p = 10^{-3}\dots 10^{-2}$ cm$^{-3}$ expected from models \citep{oph16,pog17}. The proton distribution of our empirical model implies 
$n_p =$ 0.01 cm$^{-3}$ with a radial plasma bulk speed $u_R$ = 100 km s$^{-1}$ in the heliosheath (see Section \ref{sec:IHS}). 

Outside the heliopause, a constant $n_p$ = 0.1 cm$^{-3}$ is assumed; heliospheric models
typically predict a range of $0.05\dots0.1$ cm$^{-3}$ for the region of few 100 au beyond the heliopause \citep{pog17,oph16}.
Therefore, the following ENA loss $L(E)$ is subtracted from the ENA source term in Equation~\ref{eq:ENA}:

\vspace{-7mm}

\begin{equation}
L(E) = \int_{\textup{LOS}} dl \, j_{\textup{ENA}}(E)\, n_p\, \sigma(E)\label{eq:losses}
\end{equation}

\noindent Equation~\ref{eq:losses} implies a mean free path length of 392 au for a 1 keV ENA beyond the heliopause, 
i.e., after that path length, the original ENA intensity has decreased to 1/e.

\subsection{Proton distributions}\label{sec:proton_distributions}

By far the most difficult and important task for any heliospheric ENA model is to generate the 
full intensity distribution $j_p(E)$ of all proton populations at any given place. 
Calculating the local proton density is a first step but not sufficient to predict maps of ENA intensities,
because the full angular and energy distribution of $j_p(E)$ would be required.
This is the main reason why we resorted to the simple empirical approach to predict ENA maps at this stage
of investigation.
More specifically, we defined for each of the three regions (1: inside termination shock, 2: inner heliosheath between termination
shock and heliopause, 3: beyond heliopause) the parent proton populations $j_p(E)$ 
giving rise to observable ENAs in the energy range of interest.

\subsubsection{Supersonic solar wind inside termination shock}

Inside the termination shock, we only consider neutralized protons originating from the supersonic solar wind; pick-up ions 
re-neutralized inside the termination shock are neglected so far. 
For the solar wind parameters we assume a constant $v_p$ = 440 km s$^{-1}$ ($E_p = 1.0$ keV)
representative for low heliolatitudes \citep{kha18} and $n_p(r)$ = 8 cm$^{-3}\times (\textup{1 au}/r)^2$ \citep{gos07}.
The energy distribution of the solar wind around the mean energy at a specific moment in time depends on heliocentric distance, heliolatitude, and solar activity. We approximate this energy distribution $J(E)$ with a rectangular function centered on $440\pm80$ km s$^{-1}$, based on long-term averages of Voyager 2 and New Horizons solar wind data between 11 and 31 au close to the ecliptic plane \citep{gos07,ell16}: 

\vspace{-7mm}

\begin{equation}
J(E) = 
\left(\begin{array}{cc}
\frac{F_0} {0.7 \textup{ keV}}\, , & \textup{ if } 0.7 \textup{ keV} \leq E \leq 1.4 \textup{ keV} \\
0\, , & \textup{ else.} \\
\end{array} \right)
\end{equation}

\noindent The angular distribution of the solar wind intensity is defined by the FWHM angle around the bulk direction of $\alpha_0 = 5^{\circ}$, which corresponds to a typical solar wind temperature of $1.2\times10^5$ K \citep{mar82,gos07,gal13}. 

\subsubsection{Shocked solar wind and shocked pick-up ions in the inner heliosheath}\label{sec:IHS}

Assuming that the globally distributed heliospheric ENA flux apart from the IBEX Ribbon \citep{mcc14} and the 
INCA Belt \citep{kri09} derives from the inner heliosheath \citep{gal16} and disregarding any temporal or spatial variations of that signal,
we can use the ENA measurements made with IBEX \citep{fus09,fun09}, INCA, and other ENA cameras as input for $j_p(E)$.
If $j_{\textup{HS}}$ is the intensity of the globally distributed ENA signal at 100 au in the inertial reference frame, 
the proton intensity giving rise to these measured ENAs simply is

\vspace{-7mm}

\begin{equation}
j_p(E) = \frac{j_{\textup{HS}}(E)} {(d_{\textup{HP}} - d_{\textup{TS}}) n_{\textup{H}}\, \sigma(E)}\label{eq:j_HS},
\end{equation}

\noindent provided that the proton intensities inside the heliosheath are
isotropic and constant along a radial line of sight from termination shock to heliopause. 
$d_{\textup{TS}}$ and $d_{\textup{HP}}$ denote the radial distances to the termination
shock and the heliopause, respectively.
The $j_{\textup{HS}}(E)$ in Equation~\ref{eq:j_HS} is approximated as a continuous sequence of power-laws based on ENA observations as

\vspace{-7mm}

\begin{equation}
j_{\textup{HS}}(E) = j_0 (E/E_0)^{\gamma}. \label{eq:powerlaw}
\end{equation}

\begin{figure}
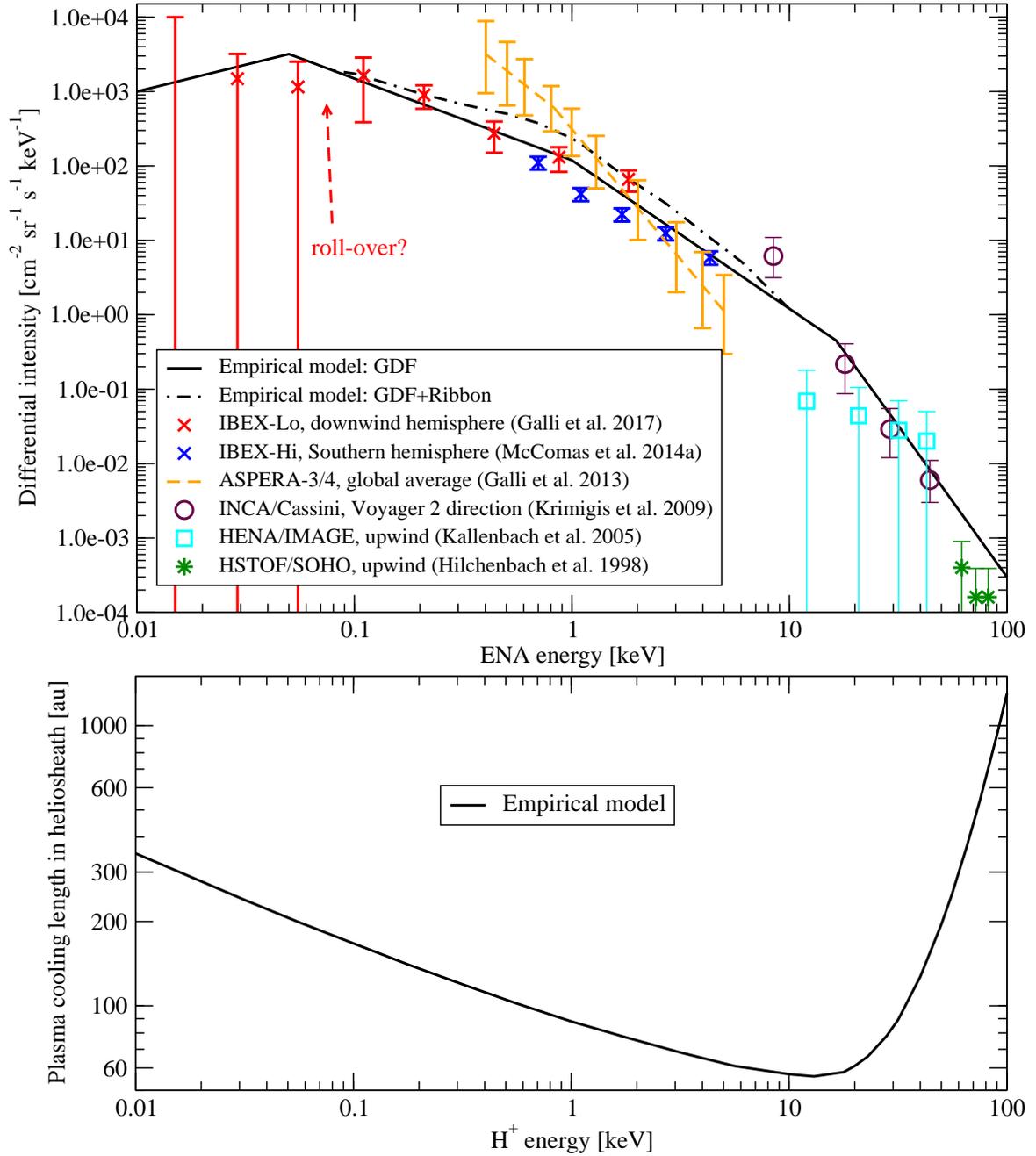

\begin{flushright}
  \includegraphics[width=0.935\textwidth]{energy_spectrum.eps}
  \includegraphics[width=0.9\textwidth,clip]{cooling_length.eps}
\end{flushright}
\caption{Energy spectra of heliospheric ENAs (upper panel). The spectra assumed for the empirical model
are plotted as black solid (globally distributed flux) and dashed-dotted line (GDF plus IBEX Ribbon ENAs),
previous observations are added as symbols. The lower panel shows the plasma cooling length in the heliosheath as a function
of the proton energy corresponding to the ENA energy in the upper panel.}\label{fig:energy_spectrum}
\end{figure}

\noindent For E = 10 to 50 eV, the power law exponent $\gamma$ is 0.72 if the heliosheath proton distribution rolls over \citep{gal16,gal17},
or $\gamma = -0.43$ if the energy spectrum gets flatter but does not roll over \citep{zir18}.
For 50 eV to 1 keV, $\gamma = -1.1$, steepening
to $\gamma = -2.0$ from 1 to 16.4 keV (based on IBEX-Lo and IBEX-Hi observations
from 2009 to 2012 at energies 50 eV to 6 keV, \citep{gal16}),
and then dropping rapidly with $\gamma = -4.0$ from 16.4 to 100 keV based on high energy ENA
measurements with INCA \citep{kri09}, HENA \citep{kal05}, and HSTOF \citep{hil98}.
This ENA energy spectrum and its observational basis are illustrated in the upper panel of Fig.~\ref{fig:energy_spectrum}. 
For this spectrum, only ENA measurements from the upwind heliosphere direction were used whenever available. 
However, for energies below 500 eV we had to rely on
downwind hemisphere measurements because of the very intense ISN signal
appearing in the upwind hemisphere in IBEX-Lo data \citep{gal14}.
The single spectrum in Fig.~\ref{fig:energy_spectrum} implies
that the heliospheric ENAs at solar wind energies and below
can be described by a spatially uniform, globally distributed flux (GDF) except for the ENA Ribbon \citep{sch14,gal16}.

The proton intensities in the inner heliosheath are modified by default by plasma loss processes. 
Based on the concept of a plasma cooling length $l_c$ \citep{sch11,gal17} and a constant 
radial plasma bulk flow of $u_R$ = 100 km s$^{-1}$, 

\vspace{-7mm}

\begin{equation}
l_c(E) = \frac{u_R} {v_{\textup{ENA}}(E)\, n_H\, \sigma(E)},\label{eq:cooling}
\end{equation}

\noindent we expect $l_c$ = 350 au, 166 au, and 57 au for 10 eV, 100 eV, and 10 keV, respectively.
The cooling length over the full energy range from 10 eV to 100 keV is plotted in the lower panel of Fig.~\ref{fig:energy_spectrum}. 
The plasma bulk flow speed $u_R$ is, in principle, not constant throughout the heliosheath \citep{zir16b}. 
However, these authors show that the variability between different models is as large as the 
modeled spatial variations. We therefore chose one global constant of $u_R$ = 100 km s$^{-1}$
between the Voyager 1 (40 km s$^{-1}$) and Voyager 2 (140 km s$^{-1}$) speed measurement \citep{sch11,glo15}.
For any $u_R$, protons around 10 keV energies happen to have the shortest cooling length, whereas $l_c$ increases
again to several 100 au for higher ENA energies. 
This implies that ENA energies below 500 eV or above 50 keV are more appropriate than intermediate energies
to image large heliosheath dimensions \citep{dem18}. Hydrogen ENAs at keV or tens of keV energy
are indicative of the plasma distribution just beyond the termination shock. 
 
We implemented plasma cooling in our empirical model via Equation~\ref{eq:cooling} 
and by assuming an exponential decrease of the local ion $j_p$ over the distance $x$ 
(radial distance from local plasma region to closest point at termination shock).
This results in a modified proton intensity of

\vspace{-7mm}

\begin{equation}
\tilde{j_p} = j_p \frac{l_{\textup{HS}}}{l_c}\frac{\exp(-x/l_c)} {(1-\exp(-l_{\textup{HS}}/l_c))}.
\label{eq:cooling_modification}
\end{equation}

\noindent The inner heliosheath thickness $l_{\textup{HS}}$ denotes the radial distance between the termination shock and the heliopause. The dimensionless normalization factor ensures that the modified proton intensity, inserted
into the basic ENA equation (Eq.~\ref{eq:ENA}), reproduces the observed input ENA intensity at 1 au. If the heliosheath thickness exceeds $10\, l_{c}$ for a specific heliosphere model 
(see Section~\ref{sec:shapes}), it is set to $10\, l_{c}$. 

Equation~\ref{eq:cooling_modification} is based on a simplification as we assume radially symmetric plasma streamlines 
in the heliosheath.
In reality, they would be curved to some extent in the flanks and towards the poles of the heliosphere, 
but the predicted curvatures also depend on the specific model (see, e.g., \citet{izm15,pog17}). 
Our empirical model does not contain plasma streamlines for the different
heliosphere shapes, and the effect would not drastically change the predicted ENA maps anyway.
Let us consider the case of largest deviations, i.e., a viewing direction towards $+z$ for the case 
of the large ellipsoid heliopause and ENA energies corresponding to $l_c = l_{\textup{HS}}$ (Eq.~\ref{eq:cooling}):
With Eq.~\ref{eq:cooling_modification} and radial streamlines, half of the total ENA intensity from the heliosheath  
along the polar line-of-sight (Eq.~\ref{eq:ENA})
is contributed by plasma between the termination shock and $0.38 \, l_{\textup{HS}}$ within the heliosheath. 
Curved plasma streamlines \citep{izm15,pog17} would narrow these ENA emissions slightly, i.e., 
the ENA half-length would reduce to a value of $(0.30...0.38) \, l_{\textup{HS}}$.   
 For viewing directions towards the nose and the downwind hemisphere, deviations of streamlines from radial symmetry are even smaller,
and for cooling lengths much shorter or longer than the heliosheath thickness, 
the effect of curved streamlines on ENA maps would also become weaker.

\subsubsection{Heliospheric ENA sources beyond the heliopause: The IBEX Ribbon}

We assume that the IBEX Ribbon of increased ENA intensities
around solar wind energies is caused by so-called secondary ENAs (see, e.g., \citet{mcc14,swa16,zir16a,mcc17,fus18,day19} 
but also see \citep{syl15} for an alternative explanation).
The secondary ENA hypothesis explains the Ribbon as due to neutralized solar wind and pick-up protons cross the termination shock and the heliopause, are re-ionized
and start gyrating around the interstellar magnetic field just beyond the heliopause 
before charge-exchanging again with the interstellar neutral hydrogen.
Again, we rely on

\vspace{-7mm}

\begin{equation}
j_p(E) = \frac{j_{\textup{Ribbon}}(E)} {d_{\textup{Ribbon}}\, n_{\textup{H}}\, \sigma(E)}\label{eq:j_Ribbon},
\end{equation}

\noindent with $j_{Ribbon}(E)=$ 250, 250, 200, 100, 35, 15, 4.5, and 0 cm$^{-2}$ sr$^{-1}$ s$^{-1}$ keV$^{-1}$ for energies 
0.1, 0.5, 0.7, 1.1, 1.7, 2.7, 4.3, and 6 keV, respectively, based on IBEX measurements \citep{gal16,mcc17}.
This energy spectrum is shown as the dashed-dotted line in Fig.~\ref{fig:energy_spectrum}.

As for the globally distributed ENA flux, temporal and spatial variations along the Ribbon
are not implemented yet. In particular, the latitudinal dependence of maximum ENA intensity
with ENA energy \citep{des19,mcc17} is not included yet.
Contrary to the globally distributed ENA flux from the inner heliosheath, these ENA contributions
are narrowly constrained in angular width and probably in thickness of their source of origin:
For a region outside the heliopause to produce Ribbon ENAs along the line-of-sight of an observer,
two conditions must be met:
First, the line-of-sight vector \textbf{r} must be nearly perpendicular to the local direction of the interstellar magnetic field \textbf{B}
\citep{mcc14,swa16}, i.e., $\mathbf{\hat{B}} * \mathbf{\hat{r}}\approx 0$ for the normalized vectors $\mathbf{\hat{B}}$ and 
$\mathbf{\hat{r}}$. The emitted ENA intensity drops off for an observer line-of-sight offset with respect to the 
$\mathbf{\hat{B}} * \mathbf{\hat{r}}=0$ surface 
as $j_p \exp(-\alpha^2/(2\alpha_{0}^2))$; the offset angle is defined as $\alpha = \lvert \arccos(\mathbf{\hat{B}} * \mathbf{\hat{r}}) - 90^{\circ}\rvert$ and the scaling factor $\alpha_0 = 5^{\circ}$ is derived from the thermal spread of the solar wind (see Section~\ref{sec:angular}). The Ribbon ENA intensity is set to zero whenever $\alpha$ exceeds
a user-defined half width of $10^{\circ}$.
Second, the location must be adjacent to the heliopause at a heliocentric distance shorter than or equal to the sum of the heliopause heliocentric distance plus the Ribbon thickness. 
For the latter we assumed a global radial thickness of $d_{\textup{Ribbon}}=$ 40 au \citep{swa16}.
The actual heliocentric distance of the Ribbon region thus depends on the heliosphere shape and viewing direction. 

To directly compare our predictions and recommended spacecraft trajectories to IBEX measurements we parameterized
the ENA Ribbon as a ring-shaped emission around a symmetry axis, i.e., the Ribbon center $\mathbf{\hat{r}_{Rb}}$. 
The local direction of the interstellar magnetic field for a given point at the heliopause thus is

\vspace{-7mm}

\begin{equation}
\mathbf{\hat{B}} = \Re_{15} \left( \begin{array}{lll} \cos(\vartheta_{Rb})\cos(\varphi_{Rb}) \\
                             \cos(\vartheta_{Rb})\sin(\varphi_{Rb})\\  \label{eq:bfield}
                              \sin(\vartheta_{Rb}) \end{array} \right).
\end{equation}

\noindent The Ribbon center $\mathbf{\hat{r}_{Rb}}$ 
is defined by $\varphi_{Rb}=-36^{\circ}$, $\vartheta_{Rb}=+35^{\circ}$
in our coordinate system, in accordance with the observed Ribbon center 
($\lambda, \beta$) = ($220.^{\circ}3,40.^{\circ}5$) \citep{day19} or ($219.^{\circ}2,39.^{\circ}9$) \citep{mcc17} in ecliptic coordinates and the upwind direction of ($\lambda=255.^{\circ}7, \beta=5.^{\circ}1$) \citep{mcc15}.
The rotation matrix $\Re_{15}$ rotates $\mathbf{\hat{B}}$ by $15^{\circ}$ away from the 
viewing direction in the ($\mathbf{\hat{r}}$, $\mathbf{\hat{r}_{Rb}}$) plane. As a result, the opening angle of the ENA Ribbon emission equals the observed radius of $75^{\circ}$ \citep{day19} instead of $90^{\circ}$.

Other potential ENA sources from the outer heliosheath are neglected in this study. For instance, the flow of interstellar plasma
along the heliopause is expected to produce a narrow fan of low-energetic ENAs around the nose of the heliosphere. Judging
from the analogy to the subsolar ENA jets observed at Mars with Mars Express \citep{fut06}, we expect 
an integral ENA intensity on the order of $10^{6}$ cm$^{-2}$ sr$^{-1}$ s$^{-1}$ between the heliopause and the bow shock
or bow wave. However, the energy of these ENAs would be concentrated
around 3 eV (corresponding to the ISN flow speed of 25 km s$^{-1}$) and thus cannot be readily detected with conventional
ENA instruments \citep{wur00}.


\subsection{Proper motion of spacecraft}

Finally, the derived integral ENA intensity at the observer may be modified by the proper motion of the observer.
For the Interstellar Probe, a radial velocity of 10 to 20 au yr$^{-1}$ is foreseen \citep{mcn18,bra19}, which translates 
to a spacecraft velocity of $u_{sc}\approx 50\dots 100$ km s$^{-1}$. This is not much lower than the speed of a 100 eV 
hydrogen ENA in the heliosphere rest frame ($v_{ENA} = 138$ km s$^{-1}$).
The resulting Compton-Getting effect increases low-energy ENA intensities from ram direction and decreases
those observed from anti-ram direction.
In our model, we assume for simplicity's sake the spacecraft is moving radially away from the Sun
with a user-defined speed of the spacecraft relative to the heliosphere between $u_{sc} =$ 0 and 100 km s$^{-1}$.
The intensity of an ENA signal following a power law (Eq.~\ref{eq:powerlaw}), observed at an angle $\varphi$ relative to the
spacecraft velocity vector, is then modified by \citep{ipa74,roe76,mcc10}:  

\vspace{-7mm}

\begin{equation}
\tilde{j}_{\textup{ENA}} = j_{\textup{ENA}} \left(1-2\cos(\varphi)\frac{u_{sc}}{v_{\textup{ENA}}} + \frac{u_{sc}^2}{v_{\textup{ENA}}^2}\right)^{(\gamma-1)}.
\label{eq:Compton-Getting}
\end{equation}


\section{Model implications for future heliospheric missions}\label{sec:applications}

The main scientific question to be answered with heliospheric ENA imaging can be formulated as follows:
How does the heliosphere interact with its galactic neighborhood and what is the 3-dimensional structure of
the heliospheric interface? The results of our empirical ENA model 
give some guidelines for an ENA imager at large heliocentric distance. 
In the following, we will mostly discuss the technical requirements of an ENA instrument
and assess the spacecraft trajectory most beneficial for heliospheric ENA imaging.
For the latter, we will focus on the geometrical aspects, i.e., parallax effects
and the opportunity to directly image plasma boundaries at large heliocentric distances. Parallax observations
have already been used in the context of IBEX observations: \citet{swa16} used parallax observations to constrain the distance of the IBEX 
Ribbon to the Earth. For a parallax baseline of $2R$ and a parallax angle $2\mu$, the distance to the object calculates to 

\vspace{-7mm}

\begin{equation}
d = \frac{R}{\tan(\mu)}
\end{equation} 

\noindent Obviously, the longer the baseline $2R$, the easier the parallax angle can be observed in ENA maps. For the IBEX
baseline of $2\times 1$ au, the observed $\mu=0.^{\circ}41$  
implied $d = 140^{+84}_{-38}$ au \citep{swa16}. 
Having a longer baseline would drastically reduce this uncertainty and would thus rule out some theories about the nature of the IBEX Ribbon. 

In addition to the geometrical aspects, observing heliospheric ENAs at large heliocentric distances
also offers benefits in terms of lower background sources (see Section~\ref{sec:snr}) and higher ENA survival probabilities. 
However, lower background rates can be achieved just by moving the spacecraft from Earth orbit to
the Lagrange 1 point of the Sun-Earth system, which is the approach for the upcoming IMAP mission \citep{mcc18}. The second benefit 
derives from the survival probabilities of heliospheric ENAs increasing with distance to the Sun: ENAs traveling to 
the inner solar system may be re-ionized and their trajectories are affected by solar radiation 
pressure and solar gravity \citep{bzo08,bzo13}. However, these effects are most relevant for low-energy ENAs with tens of eV. 
For ENAs of 100 eV or more, the survival probability to reach 1 au is at least 0.35 for any place of origin in the heliosheath \citep{gal16}.  


\subsection{Recommendations for ENA instrument specifications}\label{sec:specs}

\subsubsection{Energy range}

\begin{figure}
  \includegraphics[width=0.32\textwidth]{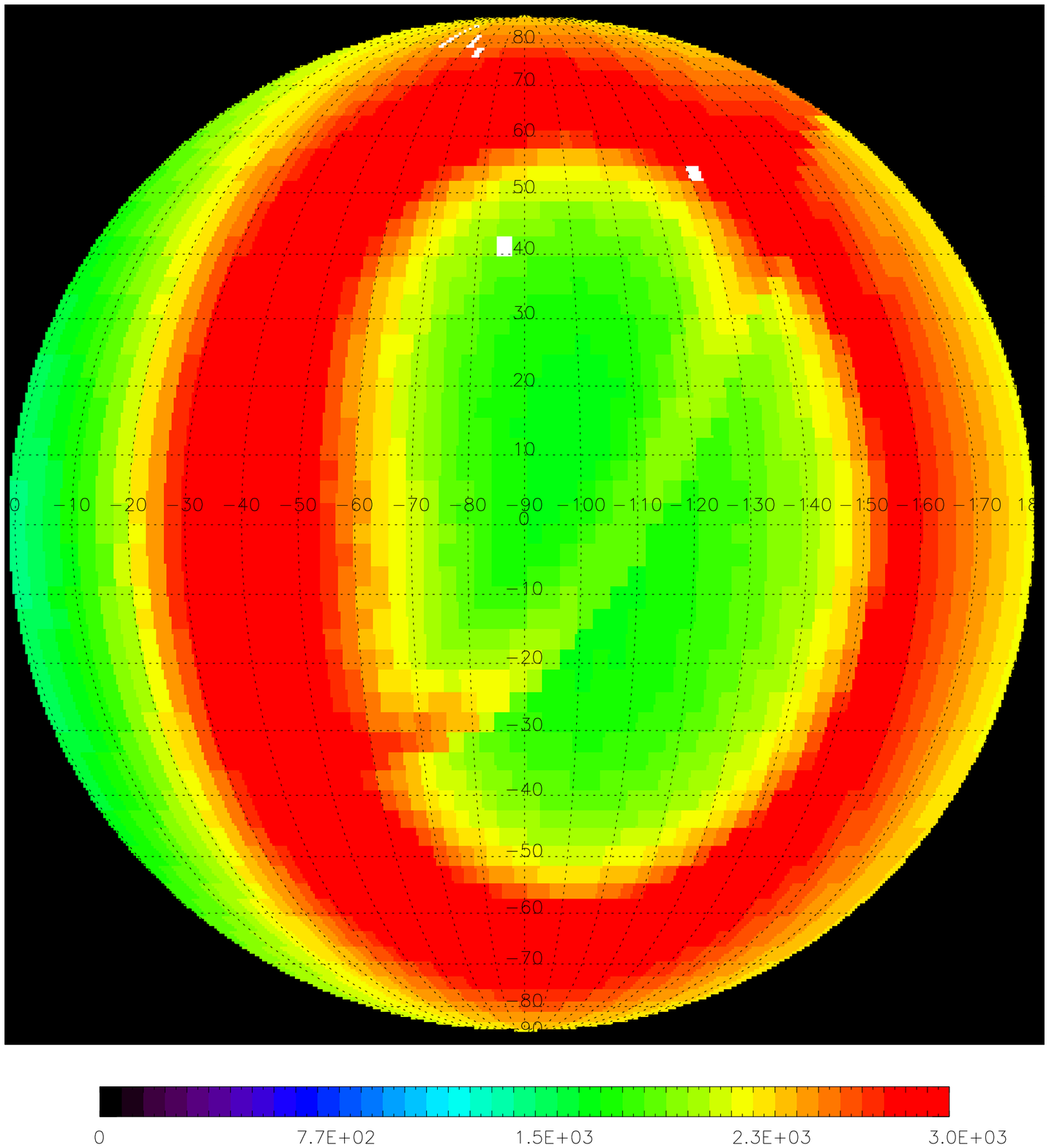}
  {\color{white} \put(-40,20){0.1 keV}}
  \includegraphics[width=0.32\textwidth]{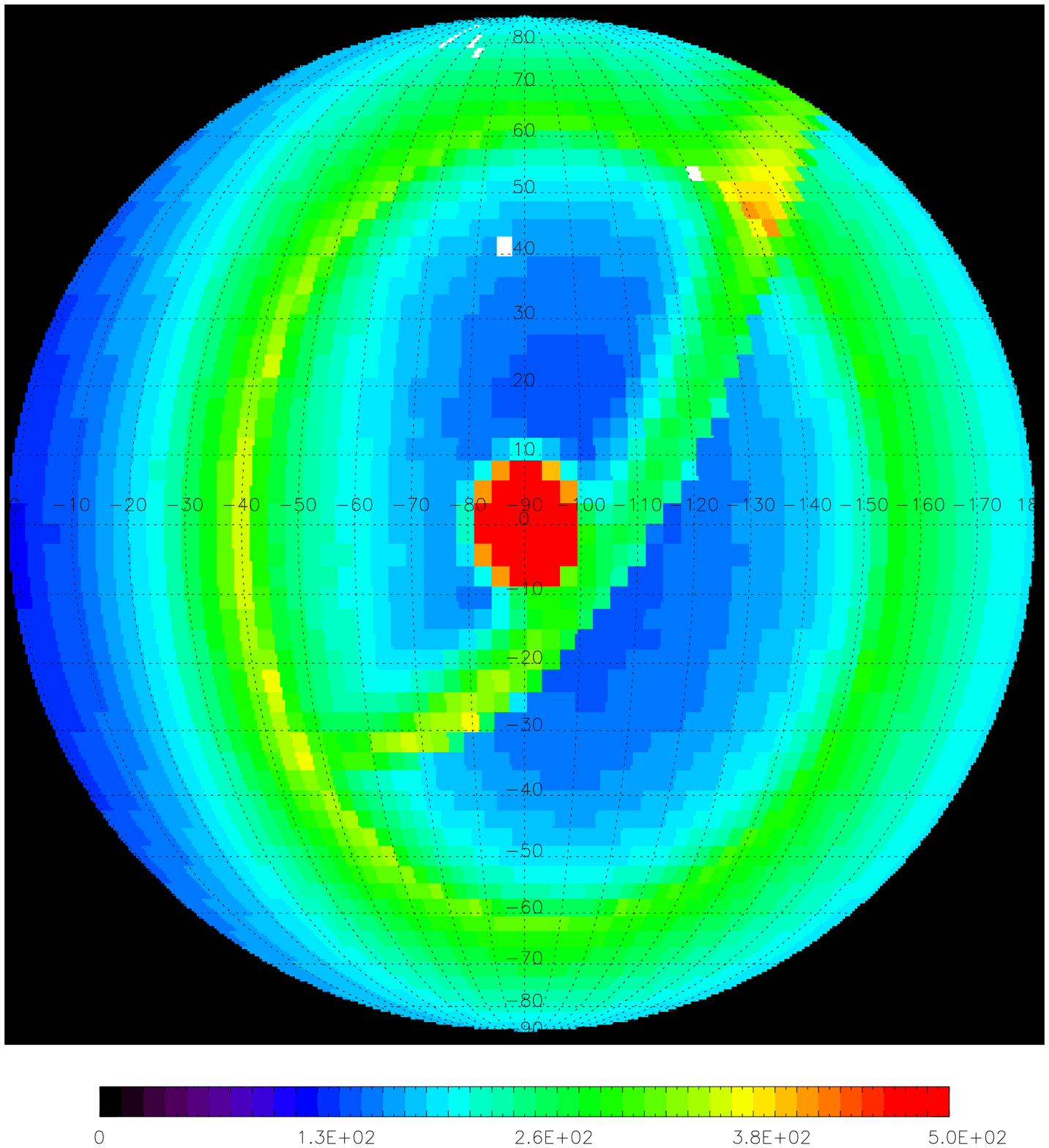}
  {\color{white} \put(-31,20){1 keV}}
  \includegraphics[width=0.32\textwidth]{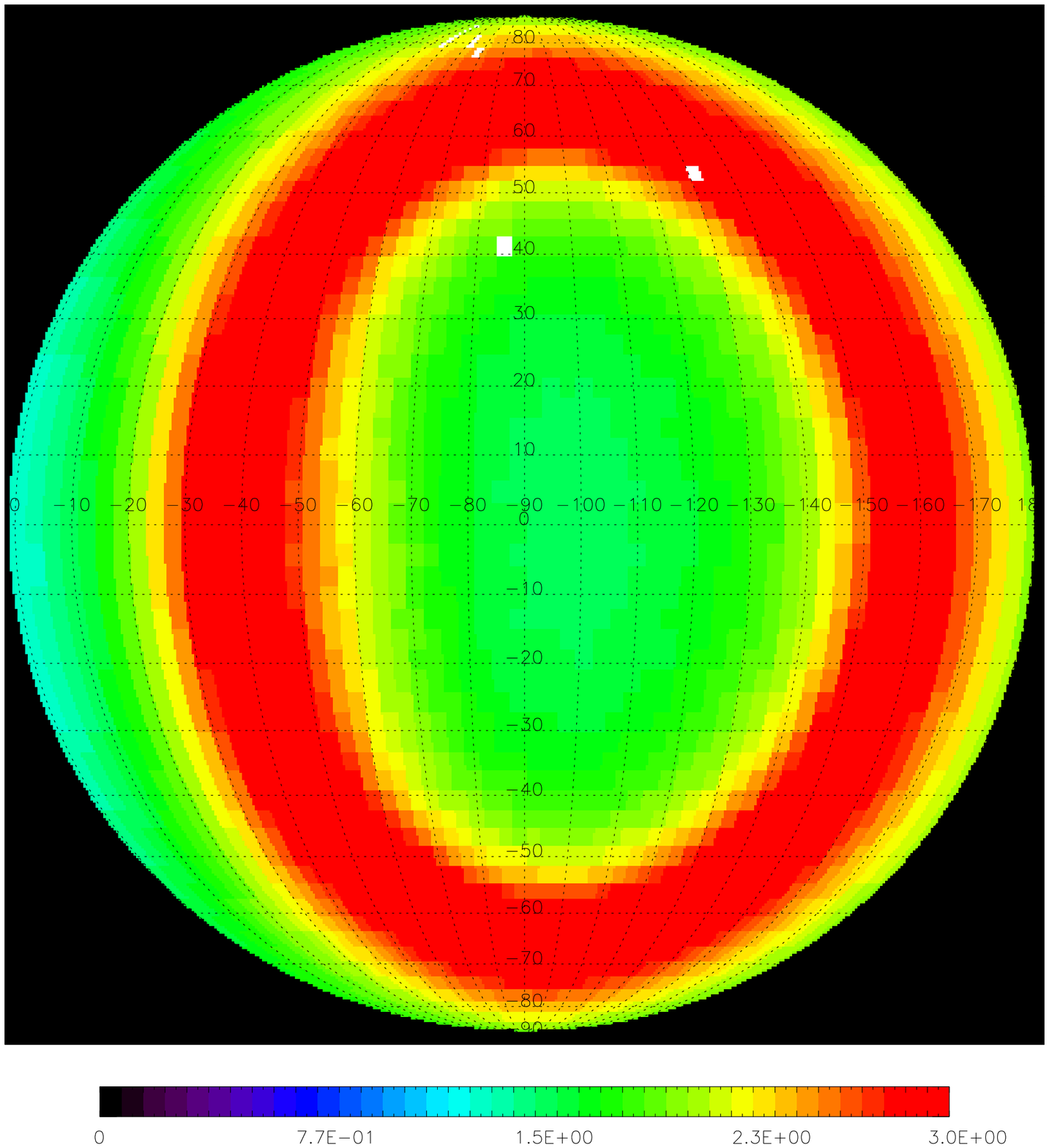}
  {\color{white} \put(-37,20){10 keV}}
\caption{ENA map predictions for an observer inside the heliosheath at 120 au heliocentric distance
for the case of a small ellipsoid heliosphere.
From left to right: ENA energies of 0.1, 1, and 10 keV (note the different intensity color scale).
The observer is located inside the ecliptic plane in the flank of the heliosheath ($\varphi=90^{\circ}$, $\vartheta=0^{\circ}$), 
looking back to the Sun in the map center.}
\label{fig:energy_comparison}
\end{figure}

Each part of the heliospheric ENA energy range provides important information about the heliosphere
and its plasma populations. On the other hand, every ENA energy range has its own observational challenges \citep{wur00}.
The general appearance of the heliosphere as seen in ENAs changes dramatically for different energies.
Figure~\ref{fig:energy_comparison} illustrates this for 0.1, 1, and 10 keV ENAs as seen from a vantage point in
the inner heliosheath at 120 au
heliocentric distance in the flank region ($\varphi = 90^{\circ}$, $\vartheta=0^{\circ}$) of a small ellipsoid heliosphere.

The protons giving rise to low energy ENAs (roughly 50 to 500 eV) dominate, by their sheer number, 
the plasma pressure in the inner heliosheath towards the flanks and downwind hemisphere \citep{gal17}.
The pressure derived for the GDF (i.e., the heliospheric ENA signal minus the IBEX Ribbon) 
is dominated by the energies from 30~eV to 500 eV for any heliospheric direction \citep{liv13}.
The physics of the heliosheath therefore can only be understood by observing ENAs in this energy range. 
One of the outstanding questions in this regard is whether the energy spectrum of the GDF rolls over around 100 eV (see Fig.~\ref{fig:energy_spectrum}). The answer to this question would help determine the importance of pick-up ions in the heliosheath and if the
GDF indeed is produced solely by plasma sources from the inner heliosheath \citep{gal17,zir18}. 
Detecting low-energy ENAs at a reasonable signal-to-noise ratio is technically more challenging
than for energies $\gg$ 1 keV. Below 300 eV, only the surface conversion technique \citep{wur00}
gives reasonable ENA detection efficiencies. That technique was successfully used by NPD/ASPERA-3\&4 \citep{bar06,bar07}
and IBEX-Lo \citep{fus09} to detect heliospheric ENAs. 
IBEX-Lo results revealed the energy spectrum of heliospheric ENAs down to roughly 100 eV,
but for lower energy the uncertainties, introduced by low count statistics and strong background
sources \citep{wur09}, made the interpretation of the results difficult \citep{fus14,gal14}.
Some of the local background sources encountered by IBEX-Lo could be avoided by an interplanetary or interstellar
ENA imager (e.g. background from Earth's magnetosphere \citep{gal16}), 
whereas other sources, in particular interstellar neutral (ISN) atoms, might persist at other places inside the heliosphere.
For a spacecraft moving away from the Sun at 20 au yr$^{-1}$ or faster,
low-energy ENAs ($\leq 50$ eV) from the anti-ram direction would not reach the spacecraft anymore, 
whereas the low-energy ENAs from the ram hemisphere would be easily detectable.

Solar wind energies (from 500 eV to several keV) are obviously vital to study the 
IBEX Ribbon in more detail at various heliocentric distances.
For these ENA energies, several detection techniques (such as conversion surfaces or conversion
foils to ionize the ENAs) are available \citep{wur00}, and the corresponding speed of these ENAs 
usually is well above the proper motion of any spacecraft. 
On the other hand, protons of solar wind energy have the shortest mean 
free path length in the heliosheath because of plasma cooling (see Equation~\ref{eq:cooling}).
Thus, they are not useful to reveal the large structures of the heliosphere outside the nose region.

High-energy ENAs above 50 keV, in contrast, have mean free path lengths of 100s of au in the heliosheath \citep{dem18}.
These ENAs represent suprathermal plasma ions and energetic particles in the heliosheath 
rather than the bulk plasma. The question remains whether ENA intensities (and their variation with time) 
at these energies reveal the actual dimension of the heliosphere \citep{dia17,sch18}. In any case, observing these
high energy ENAs is indispensable to understand acceleration processes at the plasma boundaries of the heliosphere \citep{mcc18}.

Since the detection techniques and instruments required to measure low-energy and
high energy ENAs differ strongly, at least two different ENA imagers should be foreseen to cover the
full range relevant for heliospheric ENAs. One option is to rely on three ENA imagers whereby
the low ENA imager serves a double function to also detect ISN (for ram observations close to
Earth orbit, ISN energies range from roughly 10 to 600 eV for most common ISN species). This approach
is adopted by the upcoming IMAP mission with IMAP-Lo (10 eV to 1 keV), IMAP-Hi (0.4 keV to 16 keV), 
and IMAP-Ultra (3 keV to 300 keV) \citep{mcc18}.
Another option could be to use a mass spectrometer for the ISN species and to cover the heliospheric
ENA energy spectrum with just two ENA instruments (covering the ranges
from 20 eV to few keV and from few keV to several tens of keV). This approach has the drawback that 
redundancy and energy overlap from 1 to 10 keV between the two ENA instruments might be lacking.
Having overlapping energies allows for cross-calibration between different ENA
instruments in space \citep{rei16,mcc17}. This is
very helpful as absolute calibration of ENA instruments is notoriously difficult \citep{fus09,fus12}. 

Since the whole energy range is relevant,
we will use the three ENA energies of 0.1, 1, and 10 keV as points of reference 
in the subsequent discussion of spacecraft trajectories for an ENA imager at great heliocentric 
distance (Section~\ref{sec:trajectories}).  

\subsubsection{ENA detection limits, signal-to-noise ratio, and relative uncertainty}\label{sec:snr}

The recommended ENA detection limits obviously depend on the ENA energy range of the instrument
as illustrated in Fig.~\ref{fig:energy_spectrum}. 
The signal-to-noise ratio of IBEX-Lo data, e.g., turned out to be 1 to 10 for 
energies above 100 eV, depending on viewing direction \citep{mcc17,gal17}.
The expected signal-to-noise ratio of IMAP-Lo and IMAP-Hi is higher than 50 \citep{mcc18} for ENA energies of several 100 eV.
However, the limiting factor for ENA images measured with IBEX-Lo turned out to be 
not the signal-to-noise ratio or low count statistics but rather 
strong background sources that could not be reduced by longer integration times \citep{gal14,gal16}.
These background sources can be caused by penetrating radiation, by UV light, or by ions created within the instrument
that are then post-accelerated \citep{wur00,wur09,fus05,fun09}. 

Background and noise levels are very specific to a given mission and instrument design.
General signal-to-noise or signal-to-background ratios for an interstellar
ENA imager therefore would have to invoke too many assumptions to serve as useful recommendations.
For the present study we rather ask the following question: What is the minimum significant change in ENA intensity   
if we assume an instrument
measuring the ENA intensity from a given heliosphere region at an angular resolution of a few degrees (i.e., the instrument field-of-view)
from two different observer positions?

A look at 10 years of IBEX measurements and data analysis is enlightening here: 
for IBEX-Lo, the limit for a significant intensity change is 30\% for solar wind energies and 
increases to at least 50\% for energies below 100 eV \citep{fus12,gal16}. 
IBEX-Hi has better signal-to-noise and signal-to-background ratios than IBEX-Lo 
for energies above 1 keV; the relative uncertainty for a field-of-view sized map pixel ($6^{\circ}\times6^{\circ}$)
ranges between 10 and 20\% \citep{fus14,rei16,gal17,mcc17}. If IBEX-Hi data are integrated
over a larger region in the sky (see Appendix in \citet{mcc19}) or over several months, 
the relative uncertainty can be reduced to a few \% for ENA energies above 1 keV \citep{rei16,mcc17}. 
Based on these experiences, we will assume for the subsequent section 
a critical difference of 10\% for all ENA energies. 
This is to be the minimum relative change in ENA intensity over a single map pixel
identifiable by an ENA instrument on a spacecraft at large heliocentric distance. 

\subsubsection{Angular resolution}\label{sec:angular}

The finest spatial structures in our ENA model are currently
the plasma boundaries. Other spatial structures 
necessitating higher spatial resolution may exist, 
but they are not implemented yet in our current model. Fine structures in the
IBEX Ribbon with an angular separation of a few degrees were predicted by 
\citet{gia15} and \citet{mcc18}, e.g., and such turbulence induced fine structures would change 
the observed Ribbon ENA intensities at a $2^{\circ}$ scale within a year \citep{zir19}. 
The only angular resolution inherent to our current model is 
the thermal spread of the solar wind, $\alpha_0 = 5^{\circ}$. 
Upcoming and proposed ENA imaging missions indeed aim at a similar or better
spatial resolution: for IMAP, $9^{\circ}$, $4^{\circ}$, and $2^{\circ}$ are foreseen 
depending on ENA energy \citep{mcc18}; for a LIMO-type of ENA imager at a few
au heliocentric distance, an angular resolution of $1^{\circ}$ is proposed \citep{bar19}.

If we assume, for instance, an angular resolution of $1^{\circ}$
and an ENA intensity of 100 cm$^{-2}$ sr$^{-1}$ s$^{-1}$ keV$^{-1}$ at 1 keV energy (see Fig.~\ref{fig:energy_spectrum}), 
we expect triple coincidence count rates of $10^{-4}$ to $10^{-3}$ s$^{-1}$ 
for detection efficiencies similar to IBEX-Lo \citep{fus09} and IBEX-Hi \citep{fun09} and a geometric factor reduced by 36
compared to the $6^{\circ}\times 6^{\circ}$ angular resolution for IBEX.
This implies that a few weeks would be enough to accumulate hundreds
of ENA counts and thus to obtain a statistically solid result at $1^{\circ}$ resolution at solar wind energies.
The limiting factor for angular resolution might again be the instrument-related background rates.
In the following discussions, we will use $3^{\circ}$ as the default angular resolution. 

\subsection{Which spacecraft trajectories are most interesting for heliospheric ENA imaging?}\label{sec:trajectories}

Assuming we have one or several ENA instruments onboard a spacecraft heading to heliocentric distances beyond Earth, which trajectory
is most rewarding? An actual mission concept for an interstellar probe obviously will have to balance
the science cases from many different fields such as heliosphere physics, planetary sciences, and
astronomy \citep{bra19}. Here, we restrict ourselves to the question which vantage points are useful to better understand the GDF, the Ribbon ENAs, and to determine the shape of the heliosphere. 
To this end, we created hemispherical ENA maps at 0.1, 1, and 10 keV from an assumed observer position
anywhere inside or outside the heliosphere for any of the three heliosphere
shapes. We structure the discussion of the model results the following way:

\begin{enumerate}
\item Spacecraft on a circular orbit close to the ecliptic plane at a heliocentric
distance between 2 and 10 au (Section~\ref{sec:circular}).
\item Spacecraft on a radial escape trajectory. For symmetry reasons, we only consider the quadrant covering
the Northern hemisphere from upwind to downwind direction on the port side of the heliosphere (Section~\ref{sec:radial}):

2a) Close to the ecliptic plane and headed towards the nose ($\varphi=0^{\circ}$, $\vartheta = 0^{\circ}$)  

2b) $\varphi=45^{\circ}$, $\vartheta = 0^{\circ}$

2c) $\varphi=90^{\circ}$, $\vartheta = 0^{\circ}$ (towards heliosphere flank)

2d) $\varphi=90^{\circ}$, $\vartheta = +30^{\circ}$ (above the ecliptic plane)

2e) $\varphi=180^{\circ}$, $\vartheta = 0^{\circ}$ (downwind direction, close to ecliptic plane)
\end{enumerate}

We will neglect Compton-Getting effects on ENA intensities for all trajectories to eliminate
an additional free parameter and to improve comparability between the different scenarios.
Neglecting the proper motion of the spacecraft relative to the heliospheric ENAs is
justified for a circular trajectory at large heliocentric distance. For a radial escape
trajectory, the ENA intensities seen from Sun direction would decrease and those seen from
the anti-Sun hemisphere would increase. The observed intensity would change to 0.3 and 6.7 times the inertial intensity
for the lowest considered ENA energy of 100 eV and a radial spacecraft velocity of 50 km s$^{-1}$ (see Eq.~\ref{eq:Compton-Getting}), but this would not qualitatively affect the subsequent discussion.

The Ribbon and GDF ENA sources, plasma cooling, and ENA losses outside the heliopause are included.
The subsequent figures show hemispherical images of predicted ENA intensities for an observer looking back towards the heliosphere;
all plots are centered on the Sun. The ENA maps around solar wind energy (1 keV ENAs)
are dominated by the ENA Ribbon and the very intense
direct solar wind ENA emission close to the Sun.

\subsubsection{Circular orbits and parallaxes}\label{sec:circular} 



We created hemispherical ENA maps for an observer close to the ecliptic plane at
$R=2,3,5,10$ au on a circular orbit from $\varphi=0^{\circ},90^{\circ},180^{\circ}$, and $270^{\circ}$
($\varphi = 0^{\circ}$ denoting the spacecraft position between the Sun and the nose of the heliosphere). 
The most instructive case is the comparison of the ENA maps of the upwind hemisphere as seen
from $\varphi = 0^{\circ}$ versus $\varphi = 180^{\circ}$, or seen from flank positions $\varphi = 90^{\circ}$ versus $\varphi = 270^{\circ}$. 
We always assumed the large ellipsoid as heliosphere shape here, as this makes detecting parallax effects more difficult. 
The default angular resolution of the maps 
was $3^{\circ}\times 3^{\circ}$ (see Section \ref{sec:angular}).
We then calculated the relative differences in ENA intensity from the maps obtained at 
opposite positions around the Sun. 
As motivated in Section~\ref{sec:snr}, we searched for pixels revealing 
a relative difference in predicted ENA intensity of at least 10\%.
We concentrated on 1 keV ENA energy to study the parallax effects both for the GDF and for the ENA Ribbon. 
The disadvantage of this energy is the direct solar wind signal 
(and in reality also the solar UV) blotting out the part of the maps within $20^{\circ}$ to the Sun direction.

\begin{figure}
  \includegraphics[width=0.48\textwidth]{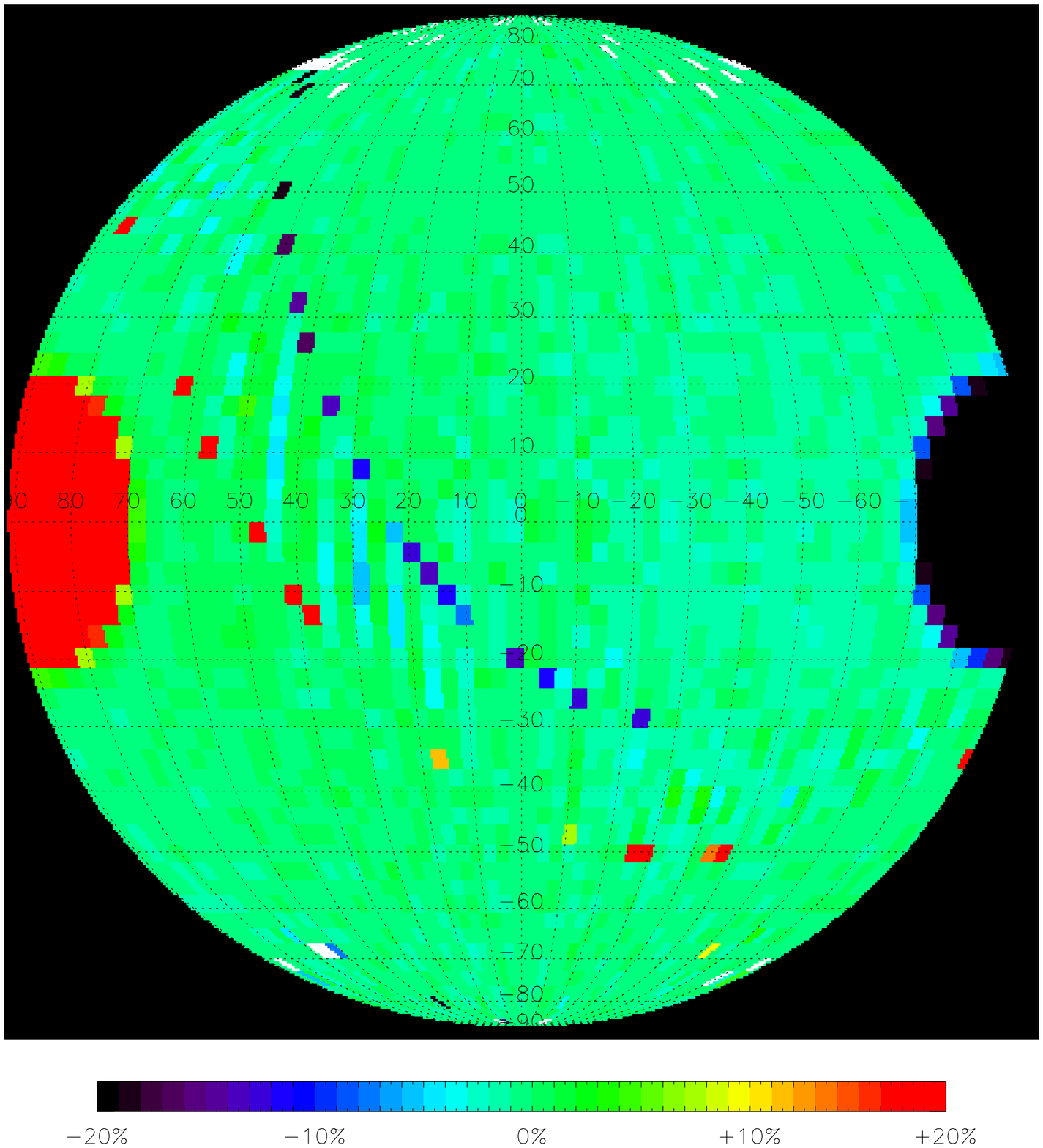}
  {\color{white} \put(-25,30){2 au}}
  \includegraphics[width=0.48\textwidth]{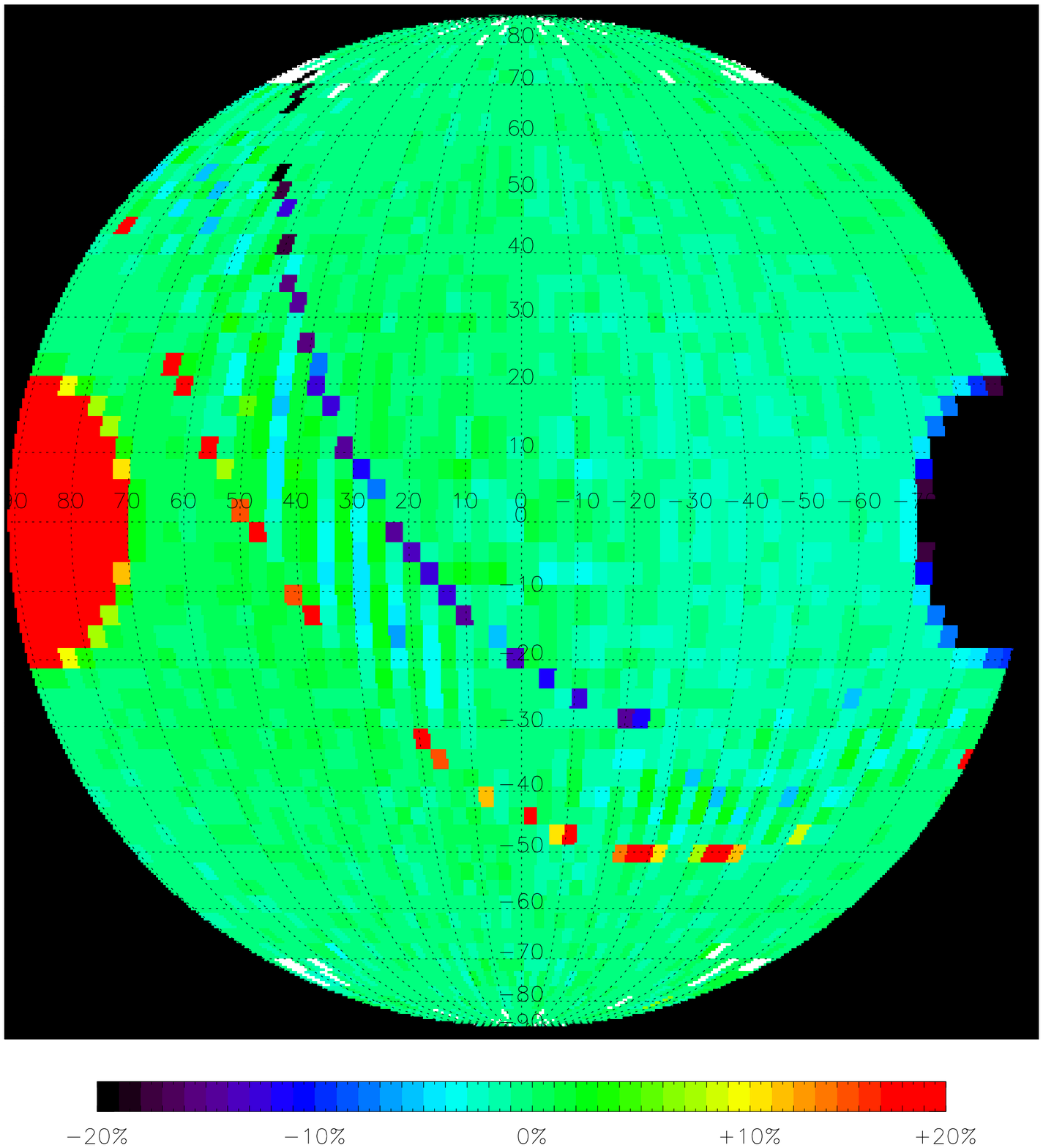}
  {\color{white} \put(-25,30){3 au}}\\
  \includegraphics[width=0.48\textwidth]{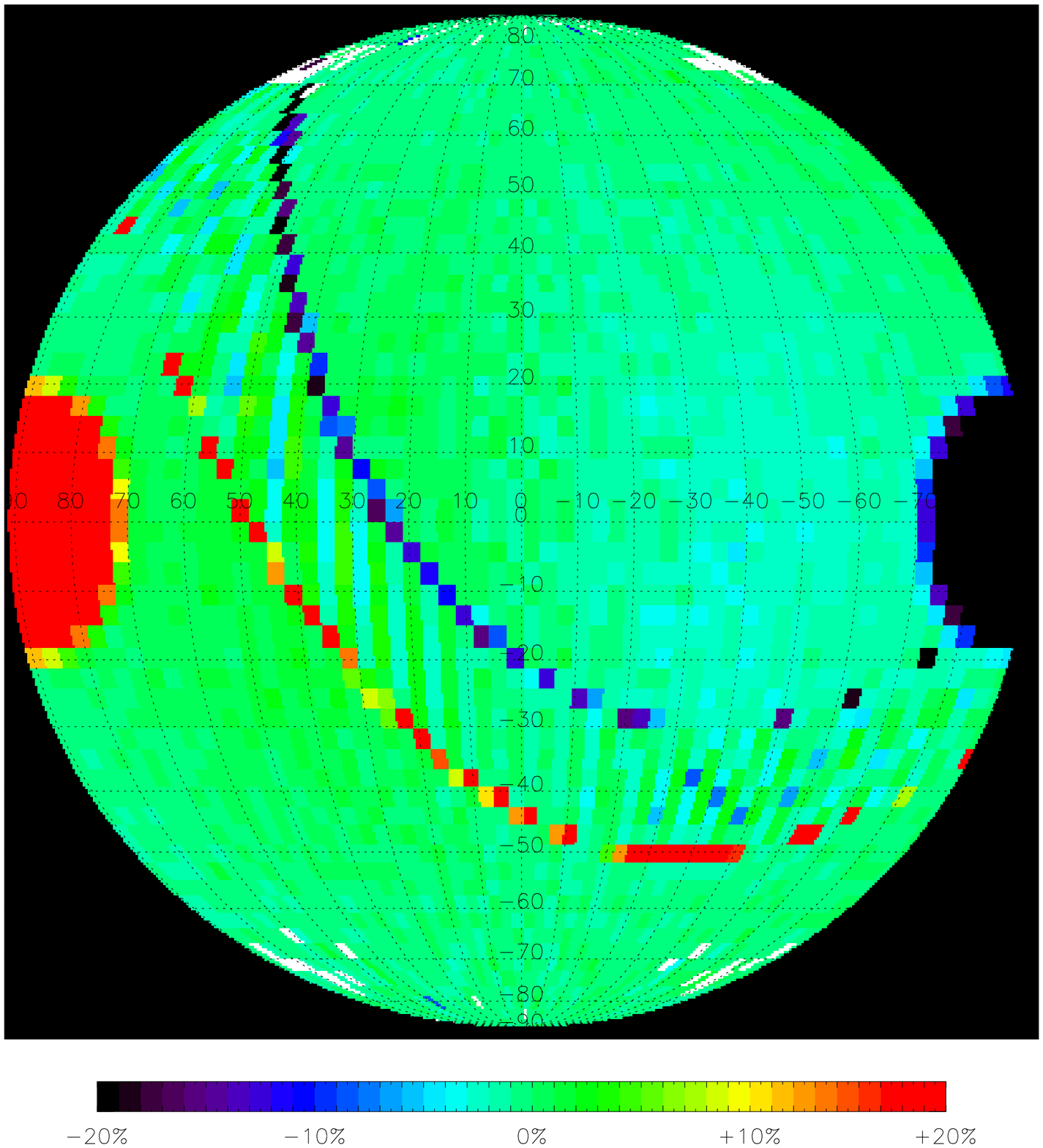}
  {\color{white} \put(-25,30){5 au}}
  \includegraphics[width=0.48\textwidth]{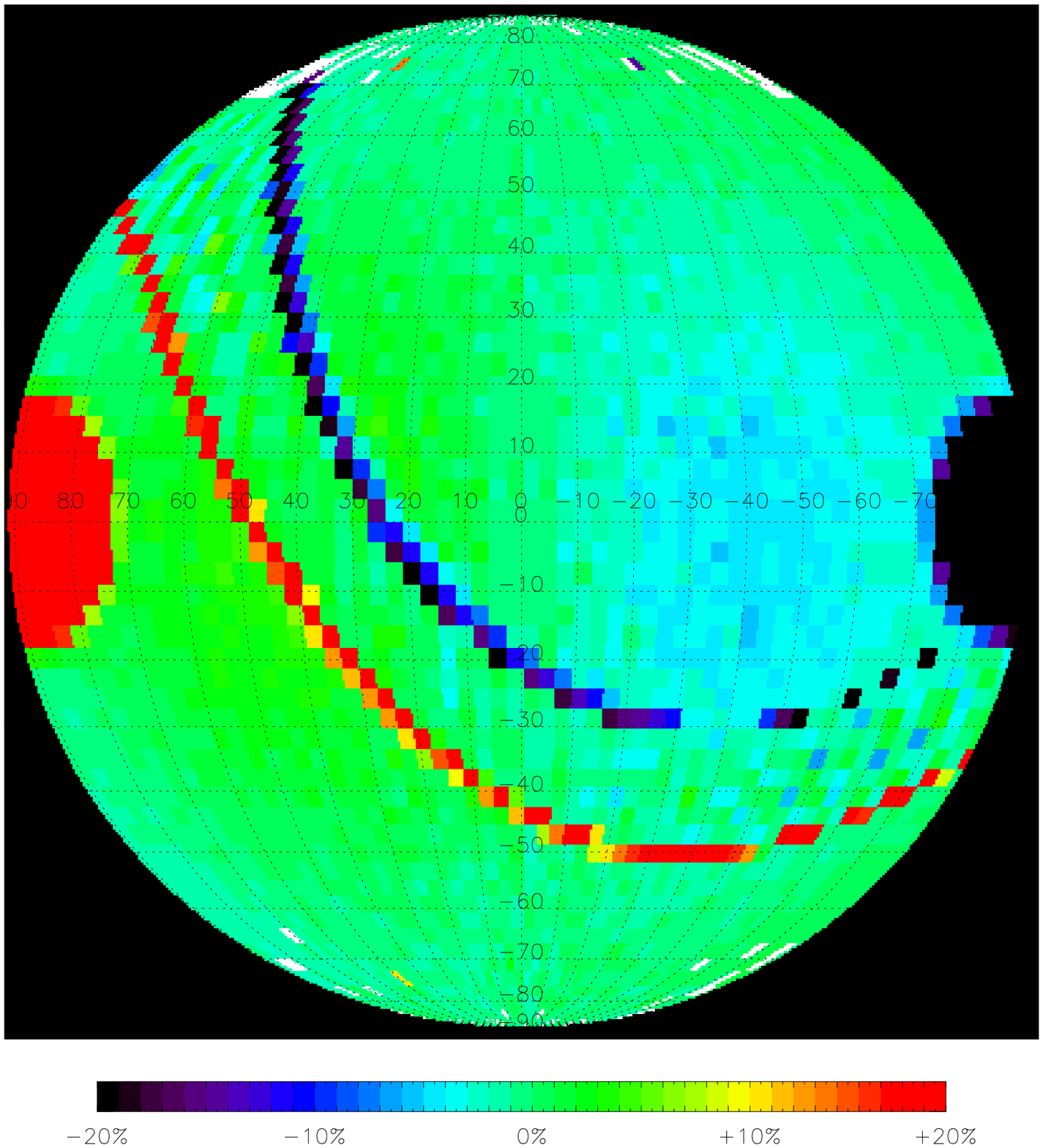}
  {\color{white} \put(-30,30){10 au}}
\caption{Relative differences of predicted ENA intensities for the upwind hemisphere
when observed from the opposite positions at $\varphi=90^{\circ}$ and $\varphi=270^{\circ}$ from 
a circular orbit at 2 au (top left) to 10 au (bottom right).
ENA energy = 1 keV, angular resolution = $3^{\circ}\times3^{\circ}$, large ellipsoid assumed as heliosphere shape.
Relative differences are detectable if pixels are blue or yellow (at least $\pm10$\% relative difference).
The Ribbon parallax feature becomes clearly visible across the hemisphere from heliocentric distances
greater than 2 au, the semicircles at the edges are introduced by the neutralized solar wind.}
\label{fig:circular_distance}
\end{figure}

\begin{figure}
  \includegraphics[width=0.48\textwidth]{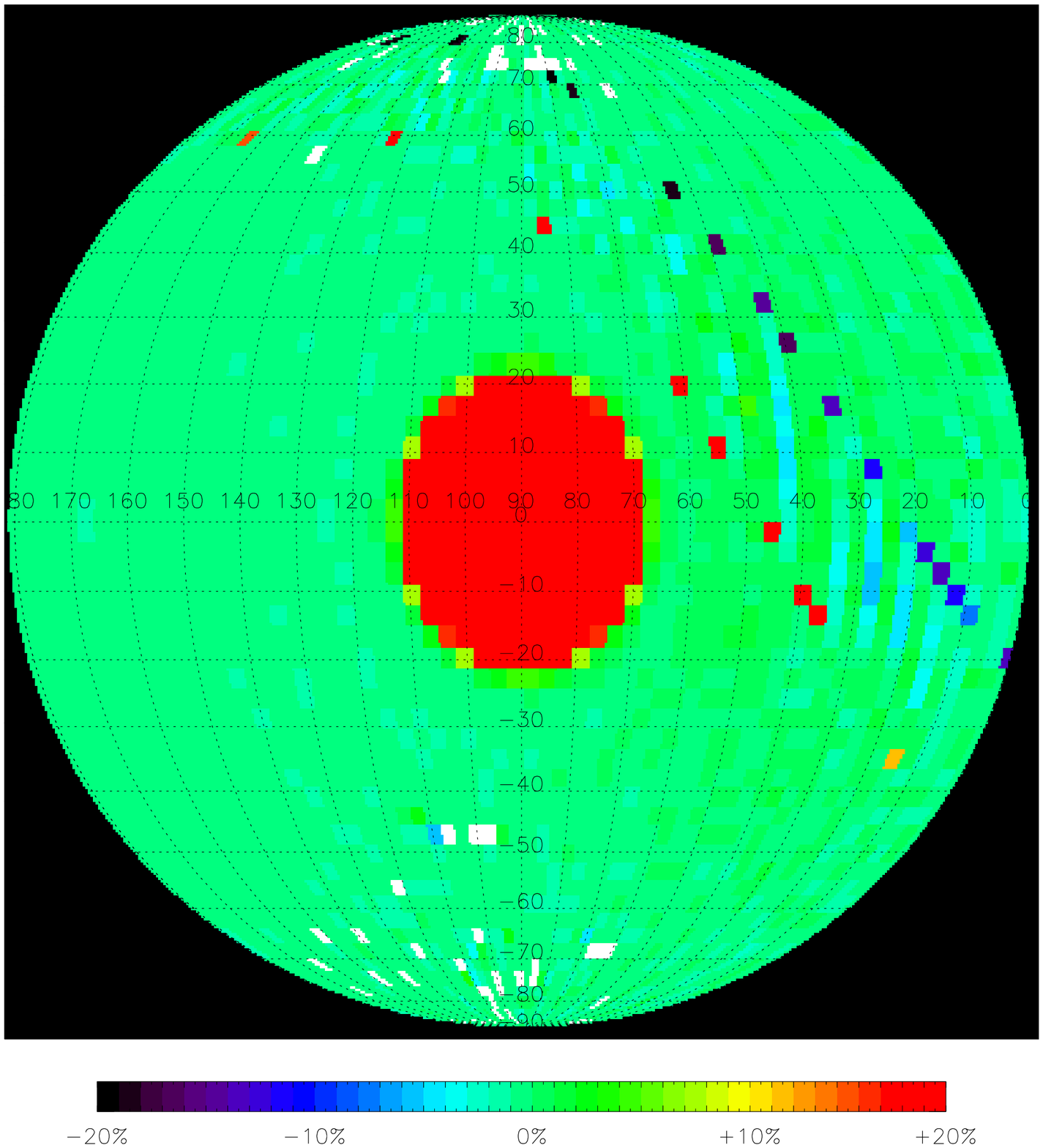}
  {\color{white} \put(-30,25){lo-res}}
  \includegraphics[width=0.48\textwidth]{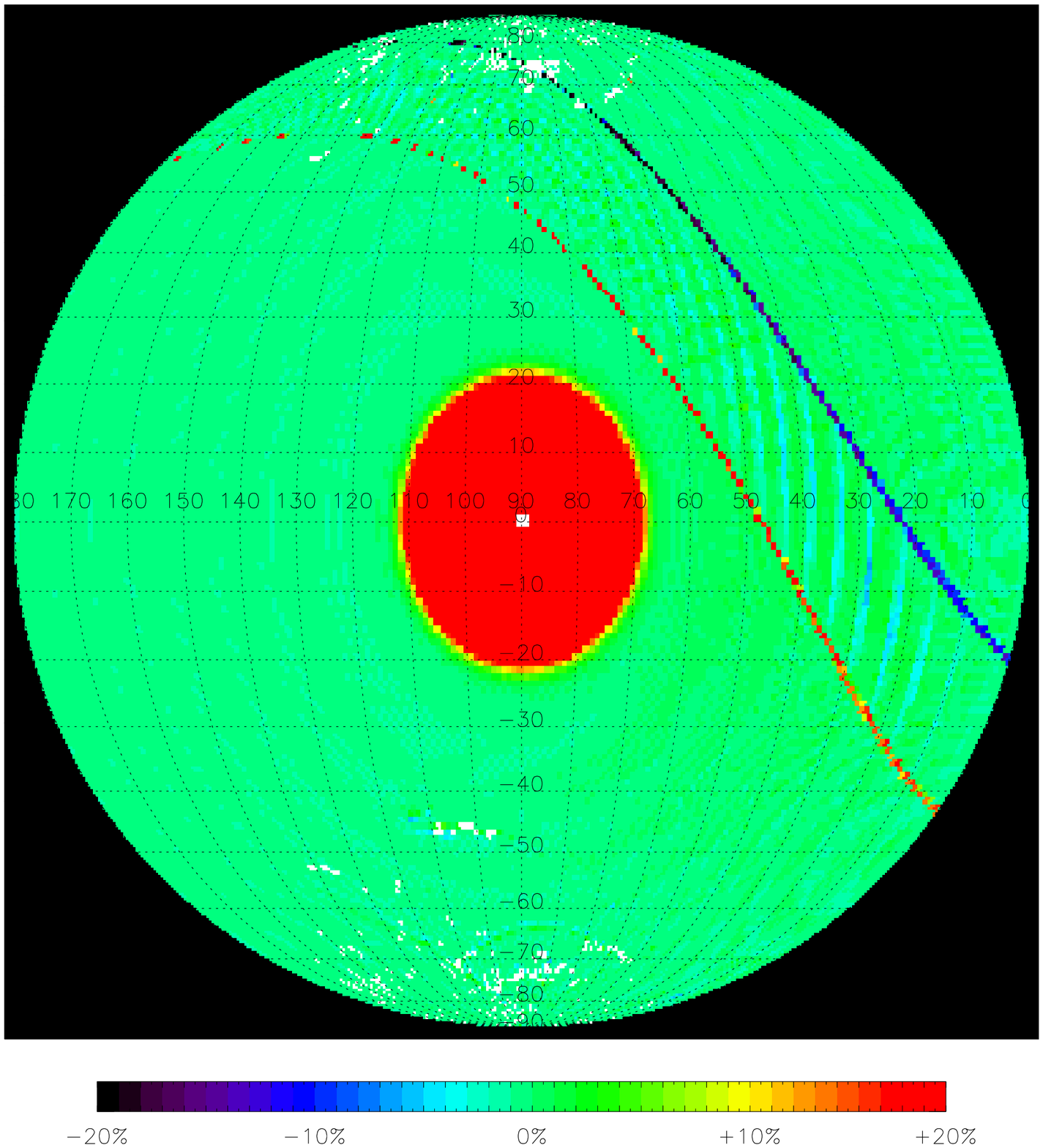}
  {\color{white} \put(-30,25){hi-res}}
\caption{Relative differences in ENA intensities at 1 keV for the flank of the heliosphere
(downwind direction at the left edge, upwind at the right edge of map)
when observed with angular resolution = $3^{\circ}\times3^{\circ}$ (left plot) versus high-resolution
of $1^{\circ}\times1^{\circ}$ (right plot). The observer vantage points used for these images
are again the opposite positions at $\varphi=90^{\circ}$ and $\varphi=270^{\circ}$ for 2 au heliocentric
distance.}
\label{fig:circular_resolution}
\end{figure}

We found that the parallax of the Ribbon induces a 10\% change in $\gg 10$ pixels in upwind hemisphere
maps if an observer baseline of $2\times3$ au or longer is assumed.
A baseline of at least $2\times5$ au is required if the downwind hemisphere is considered instead.
This holds true both for a comparison between upwind and downwind position ($\varphi=180^{\circ}$ and $\varphi=0^{\circ}$)
and for a comparison between the flank positions at $\varphi=90^{\circ}$ and $\varphi=270^{\circ}$.
Figure~\ref{fig:circular_distance} shows the relative ENA differences in the upwind hemisphere
as seen from a circular orbit at heliocentric distances increasing from 2 au (top left), to 10 au (bottom right).
For this series of plots, the maps from vantage points $\varphi=90^{\circ}$ and $\varphi=270^{\circ}$ were 
subtracted from each other. A baseline of $2\times2$ au, on the other hand, is only sufficient to spot the systematic parallax
pattern of the Ribbon if the ENA image is acquired at a $1^{\circ}\times 1^{\circ}$ 
degree resolution. This is illustrated by the comparison in Fig.~\ref{fig:circular_resolution}.

\begin{figure}
  \includegraphics[width=1.0\textwidth]{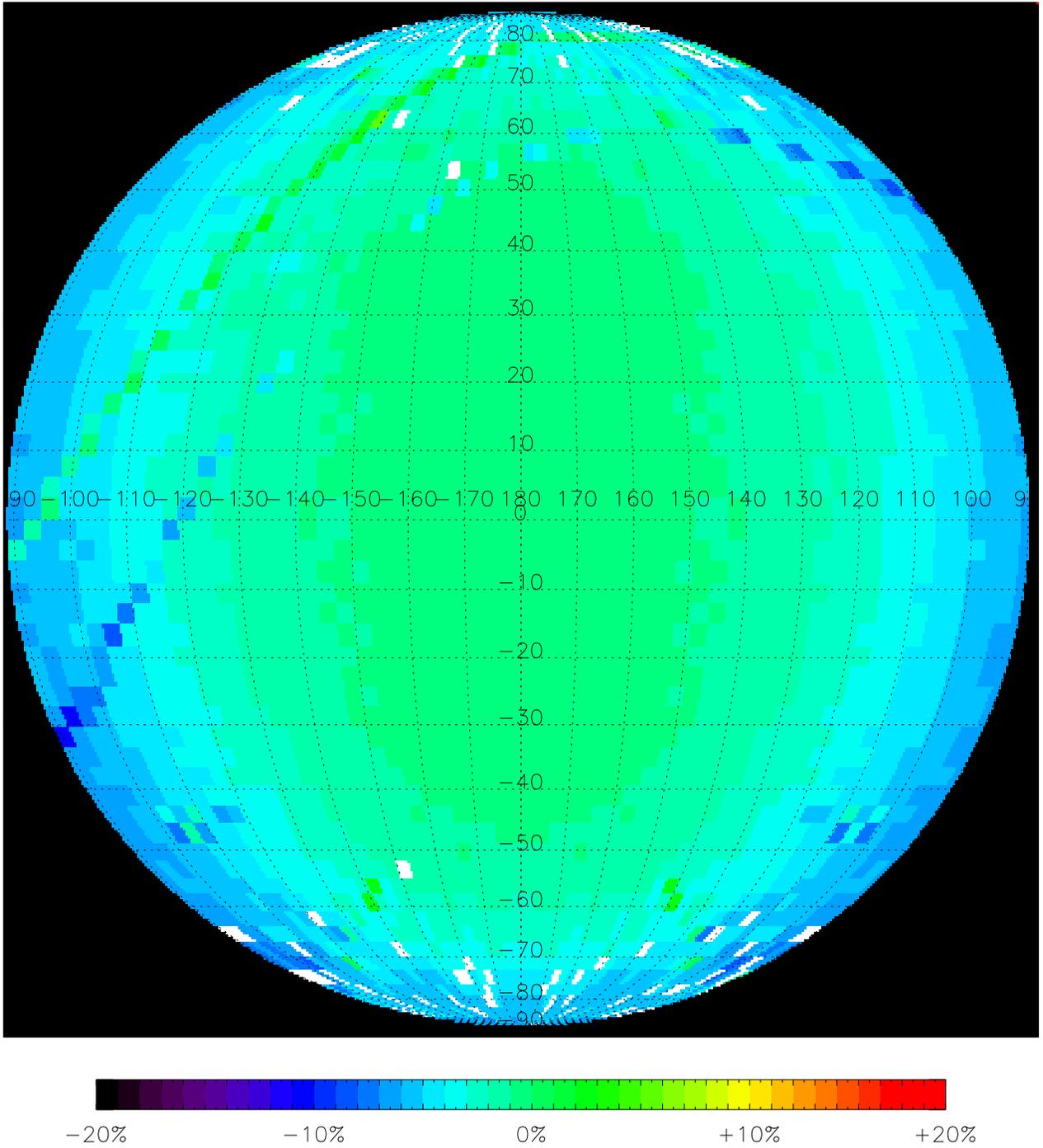}
\caption{ENA map predictions for the relative differences in ENA intensities for the downwind hemisphere
observed at 100 eV with angular resolution = $3^{\circ}\times3^{\circ}$ from vantage points
at upwind and downwind position ($\varphi=0^{\circ}$ versus $\varphi=180^{\circ}$) at 10 au heliocentric distance.}
\label{fig:circular_downwind}
\end{figure}

Apart from the Ribbon, parallax effects of the GDF can also be used -- in a much wider energy range -- 
to determine the dimensions of the inner heliosheath.
A baseline of roughly $2\times10$ au is required for systematic differences of 10\% to appear in the GDF
towards the nose of the heliosphere (bottom right plot in Fig.~\ref{fig:circular_distance}).
The relative differences in ENA intensities seen towards the flanks and poles also exceed the 10\%-limit
 for a baseline of $2\times10$ au. These differences in apparent GDF intensities are demonstrated
in Fig.~\ref{fig:circular_downwind}: it shows the differences of two ENA maps of the downwind hemisphere
observed from upwind and downwind positions $\varphi=0^{\circ}$ and $\varphi=180^{\circ}$, 10 au away from the Sun. 
This figure also shows that the GDF ENA intensity from the vast area within 
$70^{\circ}$ around the downwind direction does not change by more than few \% for a baseline of $2\times10$ au.
This holds true for any other combination of vantage points on an orbit at 10 au heliocentric distance.
The parallax effects on ENA intensities are non-detectable towards downwind either 
in the case of the small ellipsoid heliopause. 
This implies that the heliosheath dimensions in downwind direction can only be determined directly 
by ENA imaging if a spacecraft goes for heliocentric distances well beyond Saturn.

\subsubsection{Escape orbits}\label{sec:radial} 



\begin{figure}
  \includegraphics[width=0.32\textwidth]{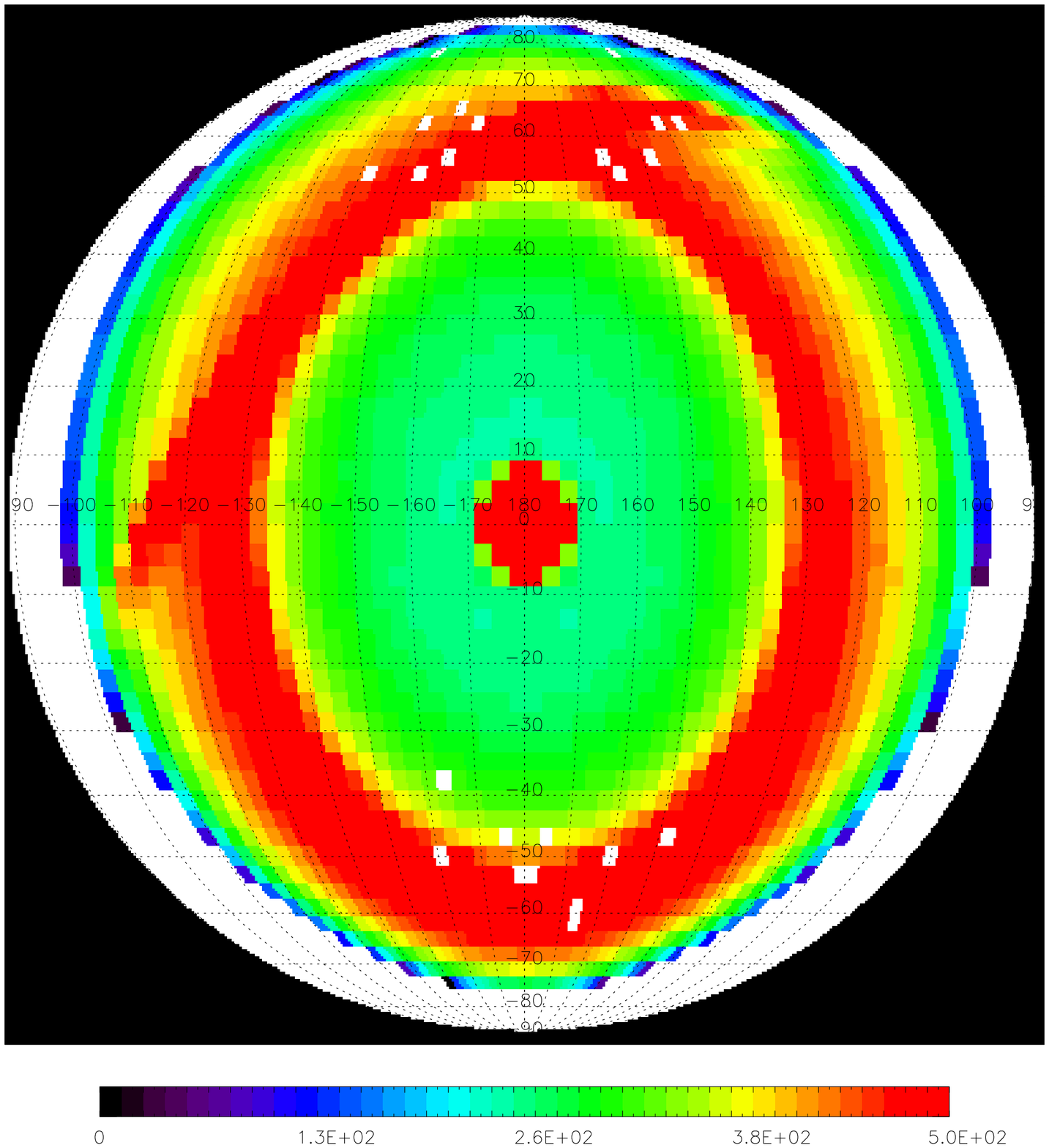}
  {\color{white} \put(-15,20){2a}}
  \includegraphics[width=0.32\textwidth]{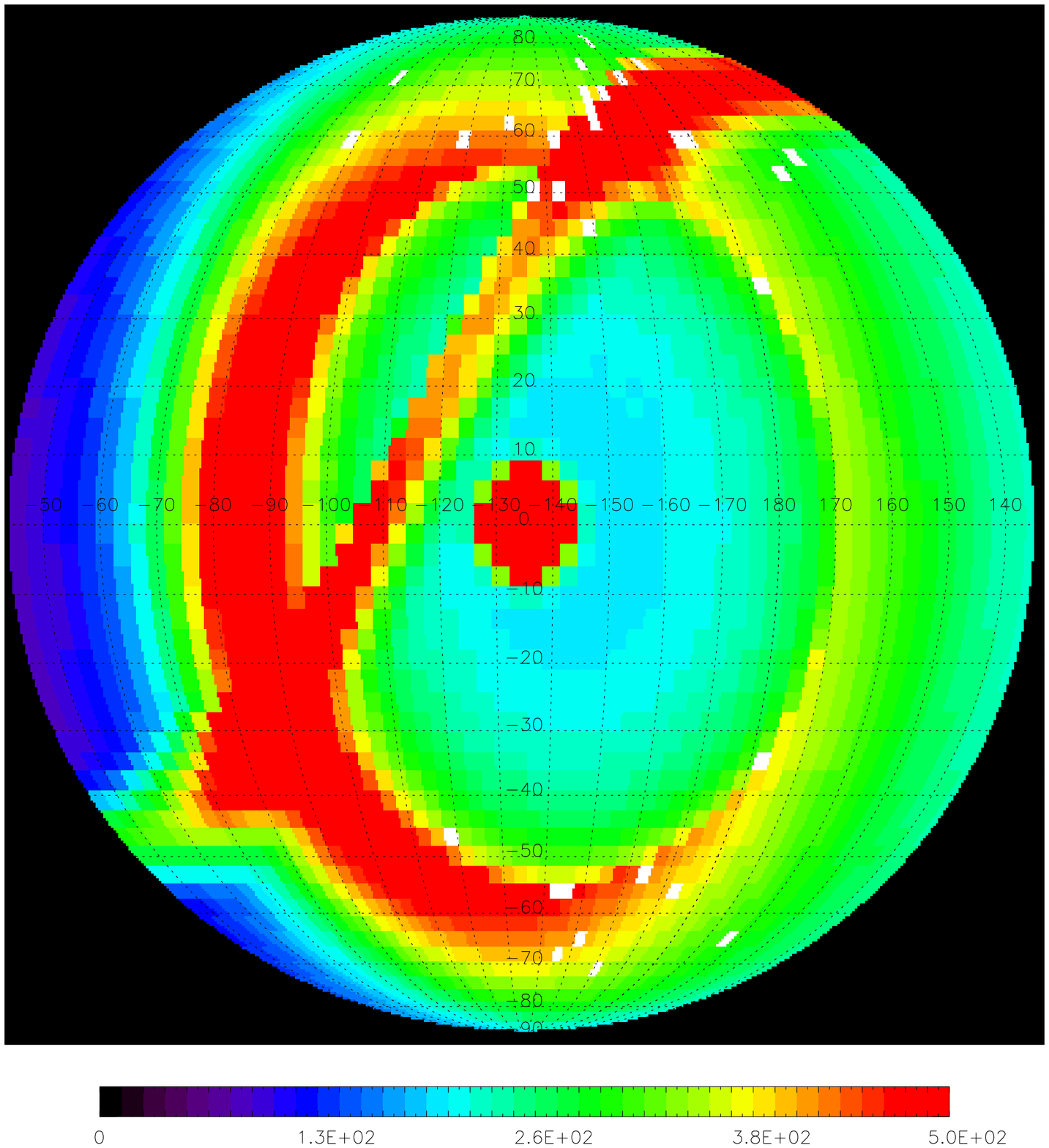}
  {\color{white} \put(-15,20){2b}}
  \includegraphics[width=0.32\textwidth]{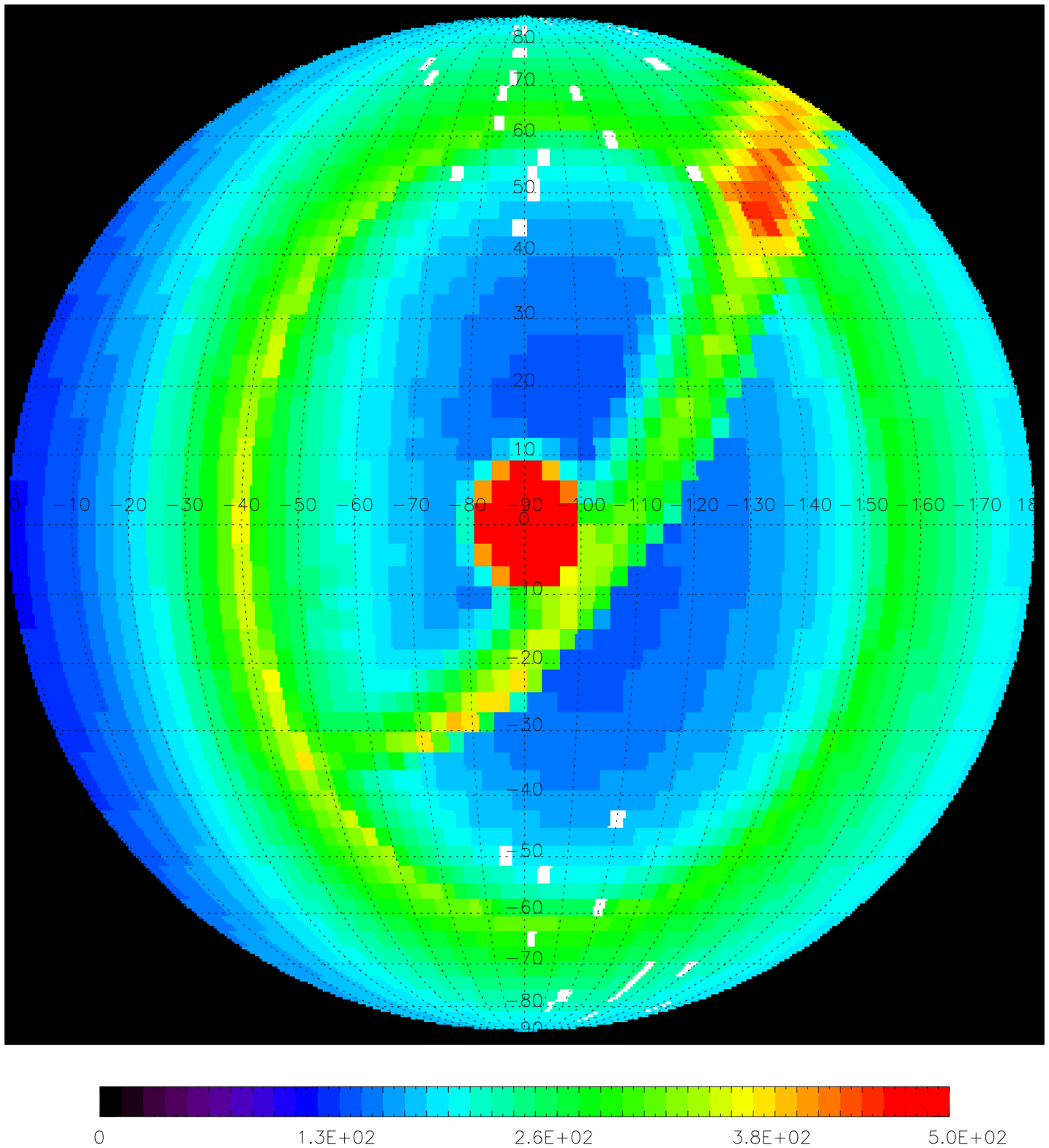}
  {\color{white} \put(-15,20){2c}}
\caption{Effect of offset between viewing angle and Ribbon center, ENA intensities predicted for a heliocentric distance
of 120 au at 1 keV for a large ellipsoid heliosphere. Left:
observer placed in front the nose of the heliosphere (trajectory 2a: $\varphi=0^{\circ}$, $\vartheta=0^{\circ}$), middle:
observer placed at (trajectory 2b: $\varphi=45^{\circ}$, $\vartheta=0^{\circ}$). Right: observer placed in the flank of the heliosheath with a $90^{\circ}$ angle between the Sun-spacecraft and the Sun-upstream line (trajectory 2c: $\varphi=90^{\circ}$, $\vartheta=0^{\circ}$).}
\label{fig:flank_minimum}
\end{figure}

The optimum trajectory of a future interstellar ENA imager must reach a compromise 
between reaching the heliopause within a reasonable time, imaging the hitherto unknown downwind regions, 
and being able to separate the Ribbon ENAs from the GDF from the heliosheath. 
The trajectory should offer viewing angles rather perpendicular to both 
the symmetry axis of the heliosphere and to the Sun-Ribbon center line.
For some longitudes, the maximum Ribbon ENA emissions will strongly overlap with those from the GDF around
the plasma boundaries, which would complicate interpretation of both ENA sources. This is illustrated by the comparison
of the ENA maps (Fig.~\ref{fig:flank_minimum}) predicted for trajectories 2a, 2b, and 2c (from left to right)
for 120 au heliocentric distance and 1 keV ENA energy. 
Trajectory 2a means heading directly towards the nose ($\lambda \approx 256^{\circ}$, $\beta \approx 5^{\circ}$ in ecliptic coordinates), 
Trajectory 2b corresponds to a path towards the Ribbon region close to the ecliptic plane 
($\lambda \approx 301^{\circ}$), 
half way between nose and flank on the port side of the heliosphere, with a $45^{\circ}$ offset to the upwind direction and $81^{\circ}$ to the Ribbon center. 
Trajectory 2c implies the flank on the port side ($\lambda \approx 346^{\circ}$), 
$90^{\circ}$ away from the nose of the heliosphere. 

\begin{figure}
  \includegraphics[width=0.32\textwidth]{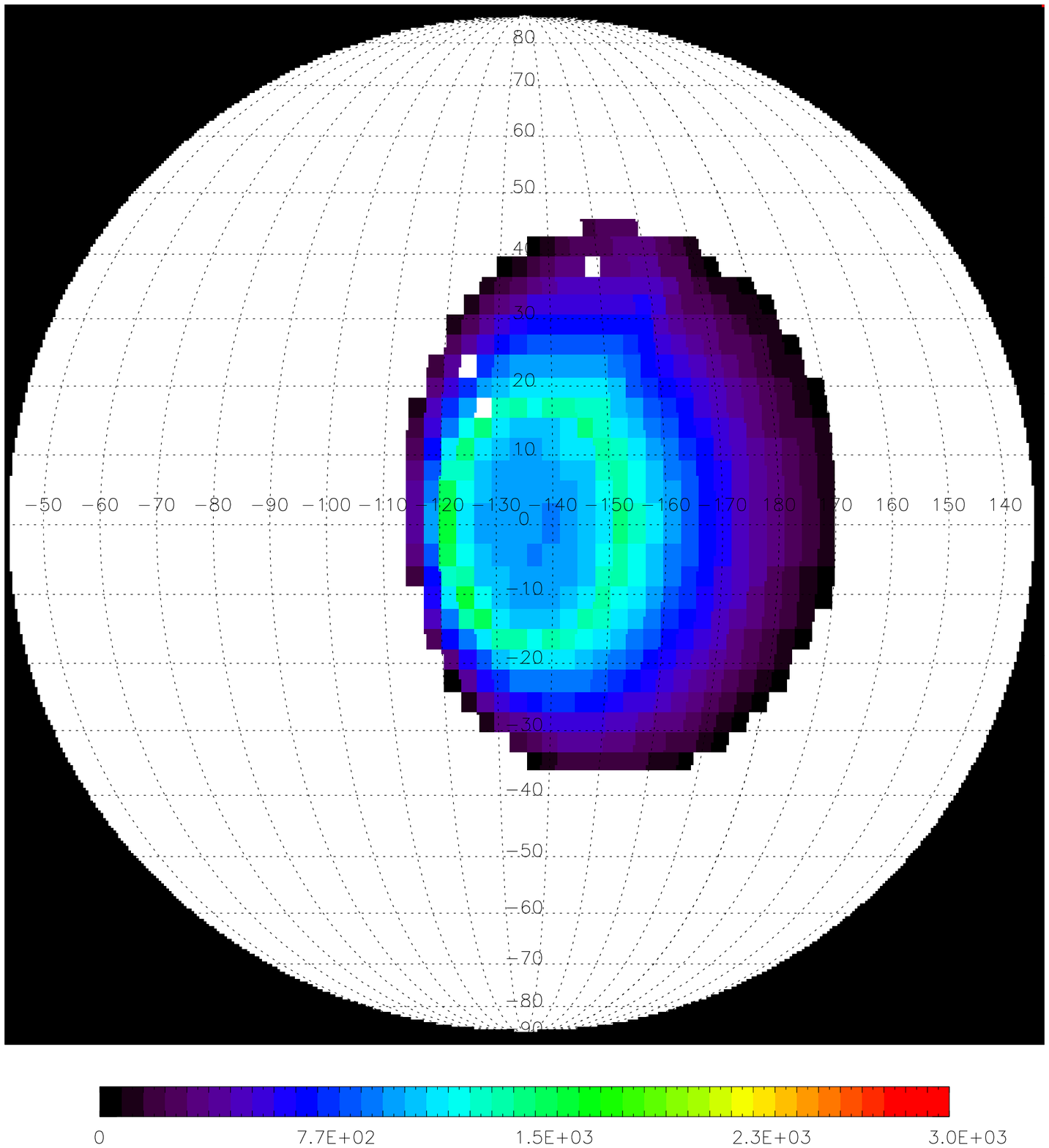}
  {\color{white} \put(-35,20){Sm,2b}}
  \includegraphics[width=0.32\textwidth]{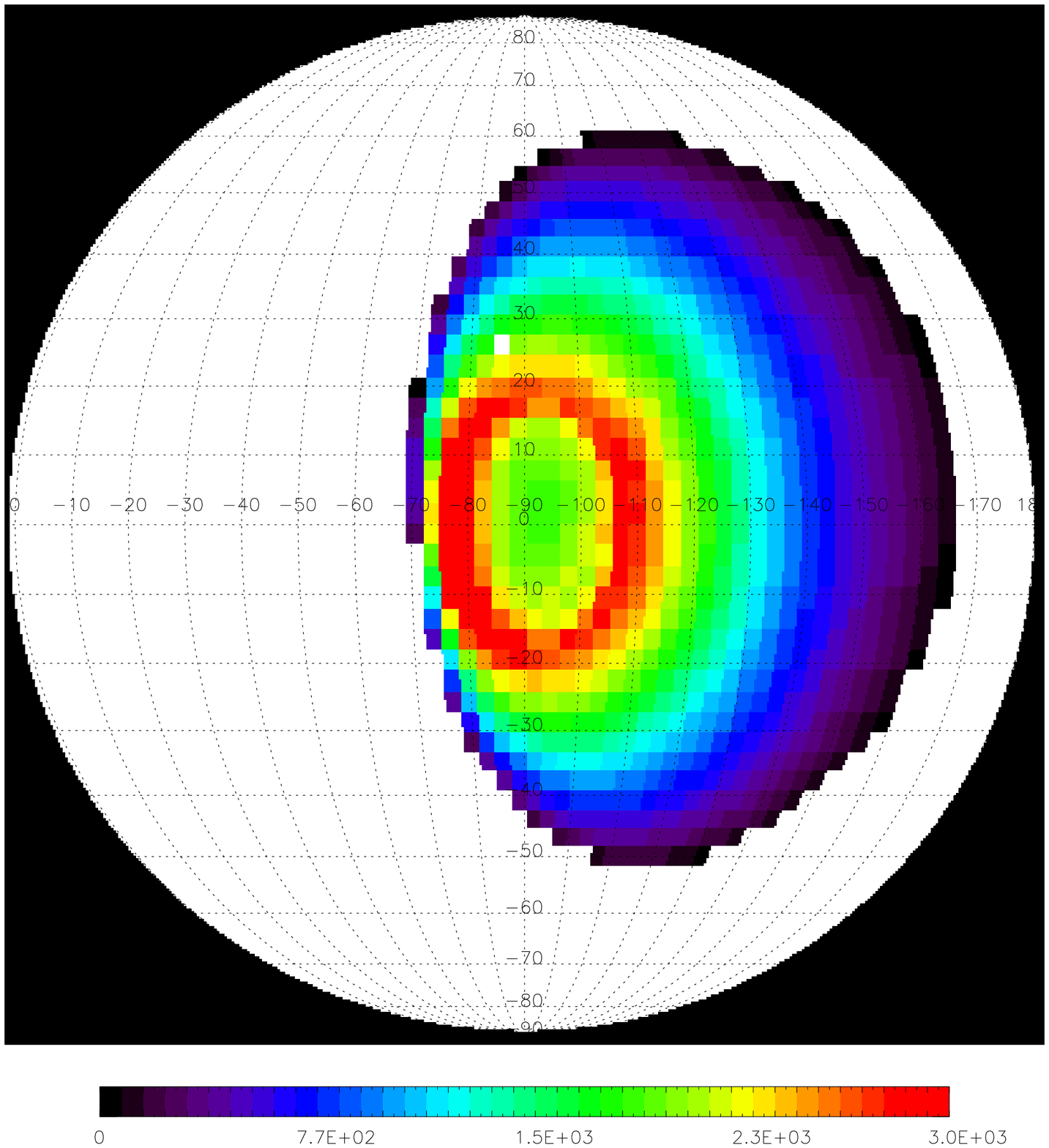}
  {\color{white} \put(-35,20){Sm,2c}}
  \includegraphics[width=0.32\textwidth]{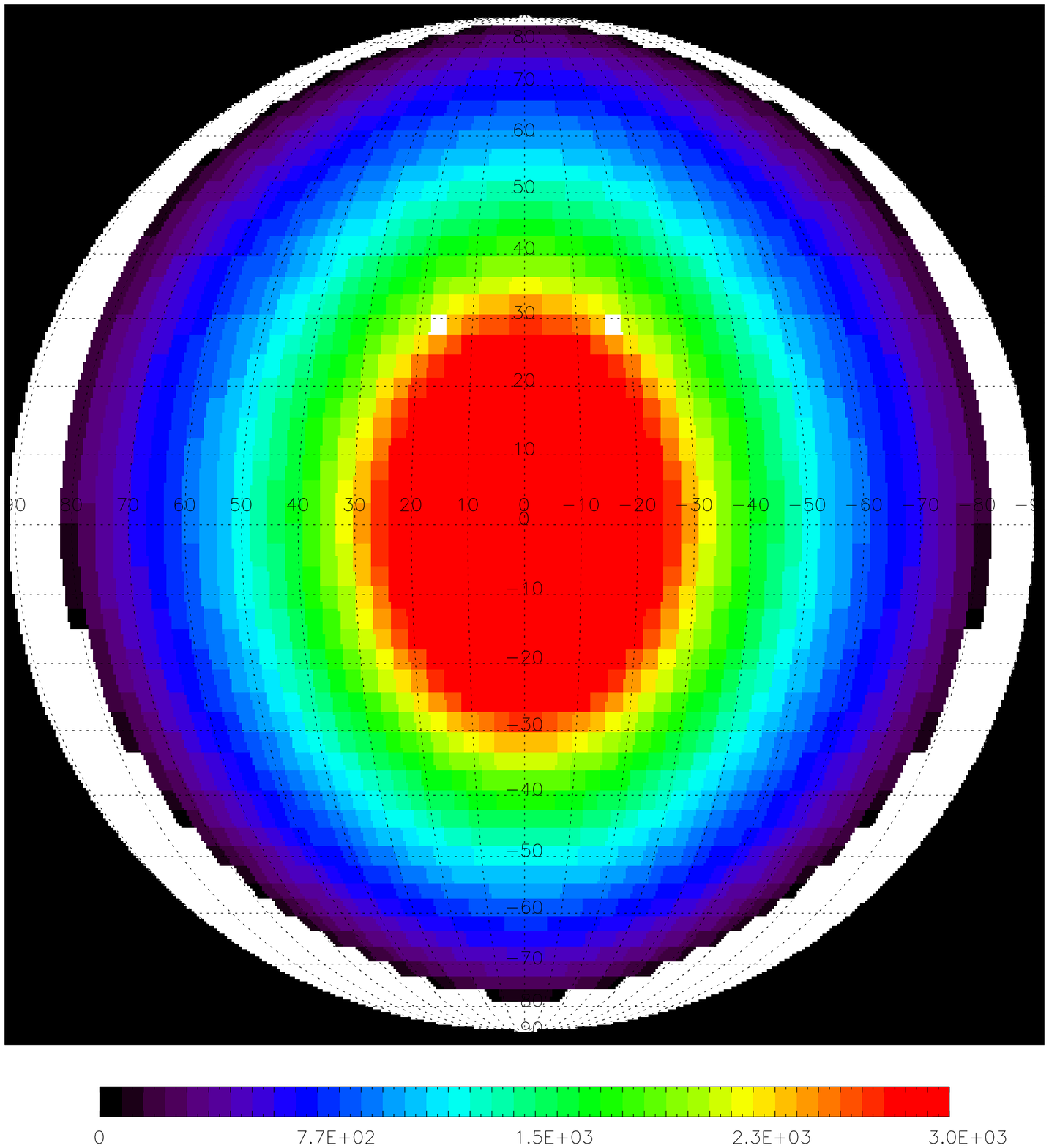}
  {\color{white} \put(-35,20){Sm,2e}}\\
  \includegraphics[width=0.32\textwidth]{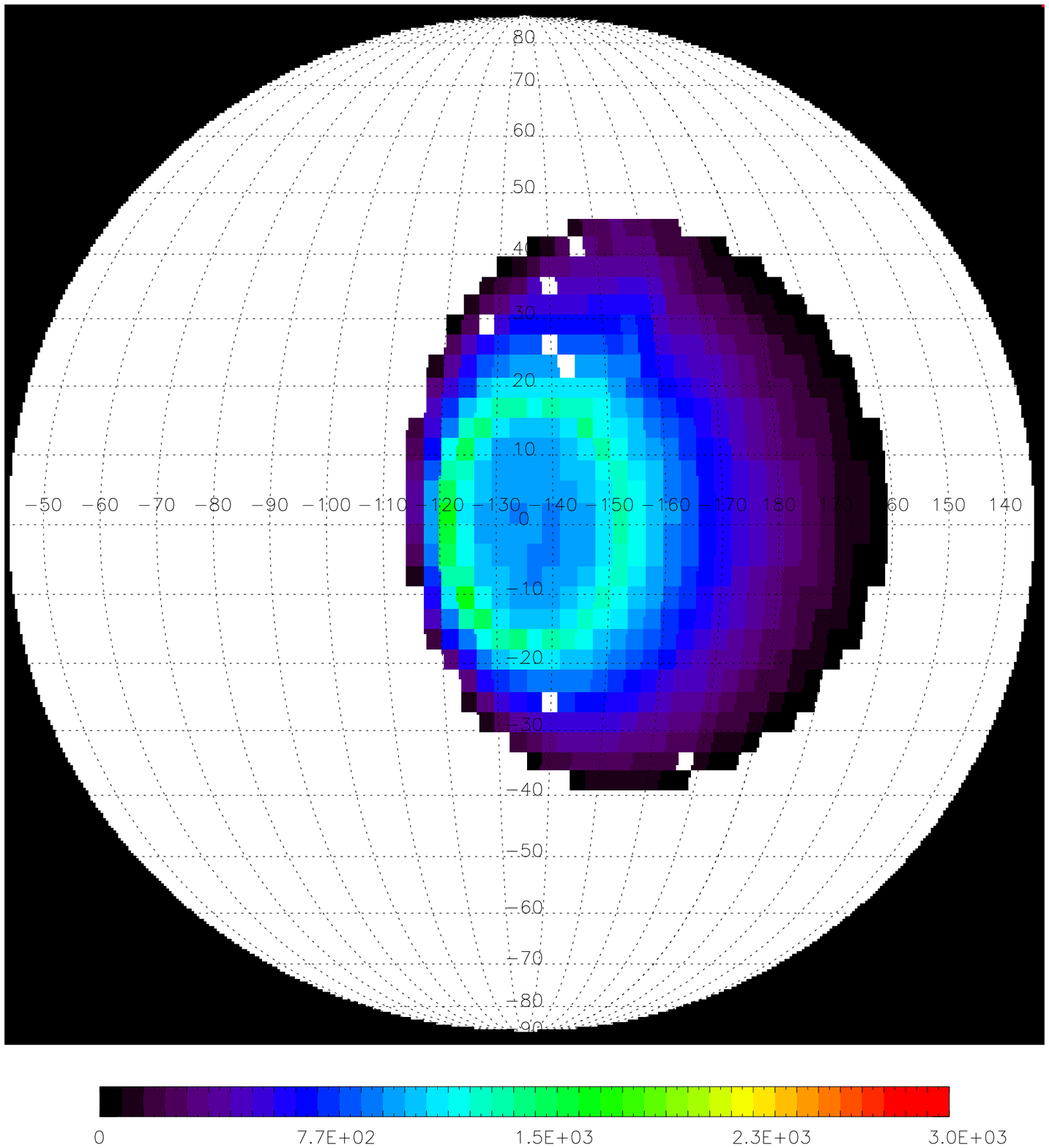}
  {\color{white} \put(-31,20){La,2b}}
  \includegraphics[width=0.32\textwidth]{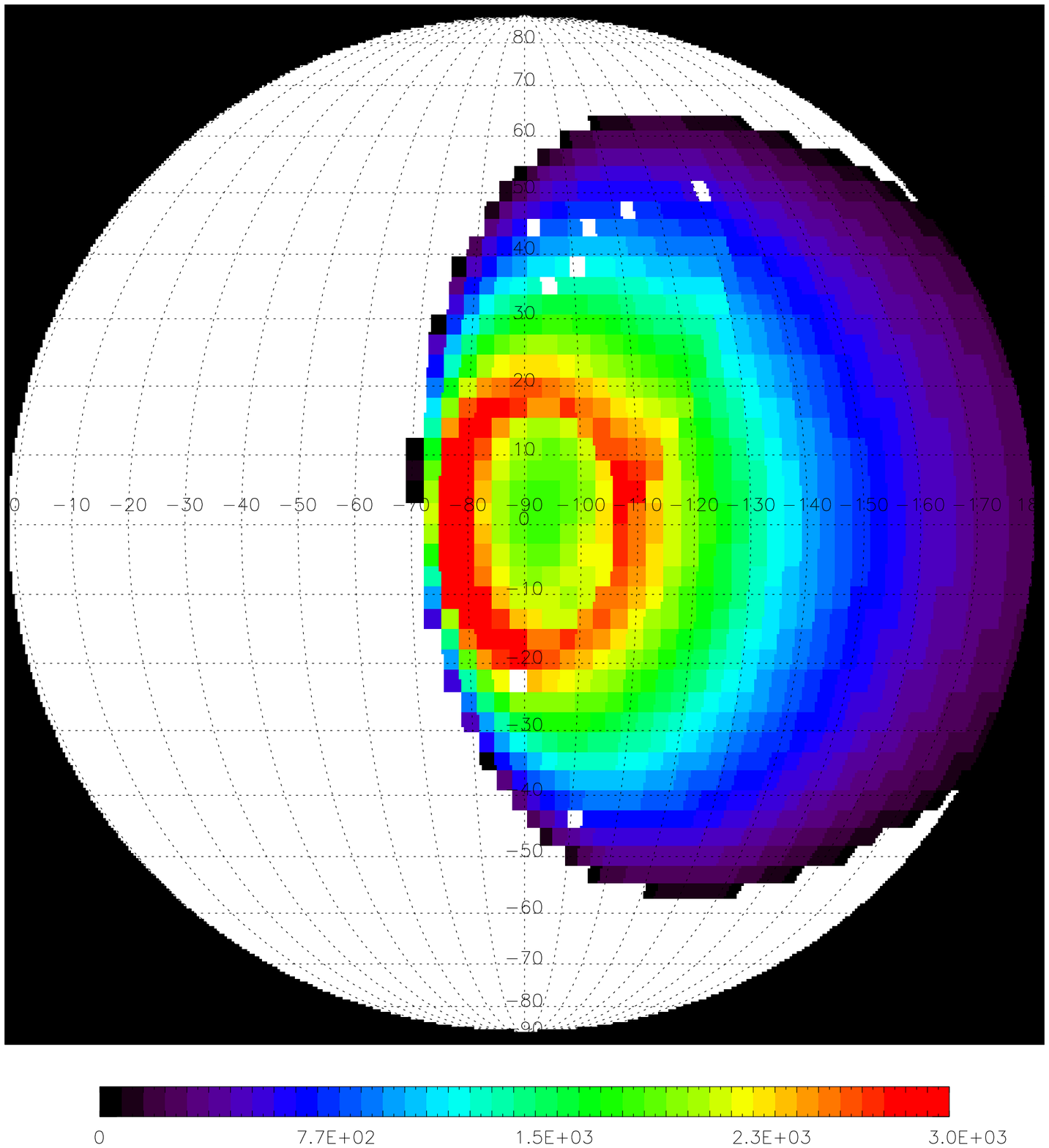}
  {\color{white} \put(-30,20){La,2c}}
  \includegraphics[width=0.32\textwidth]{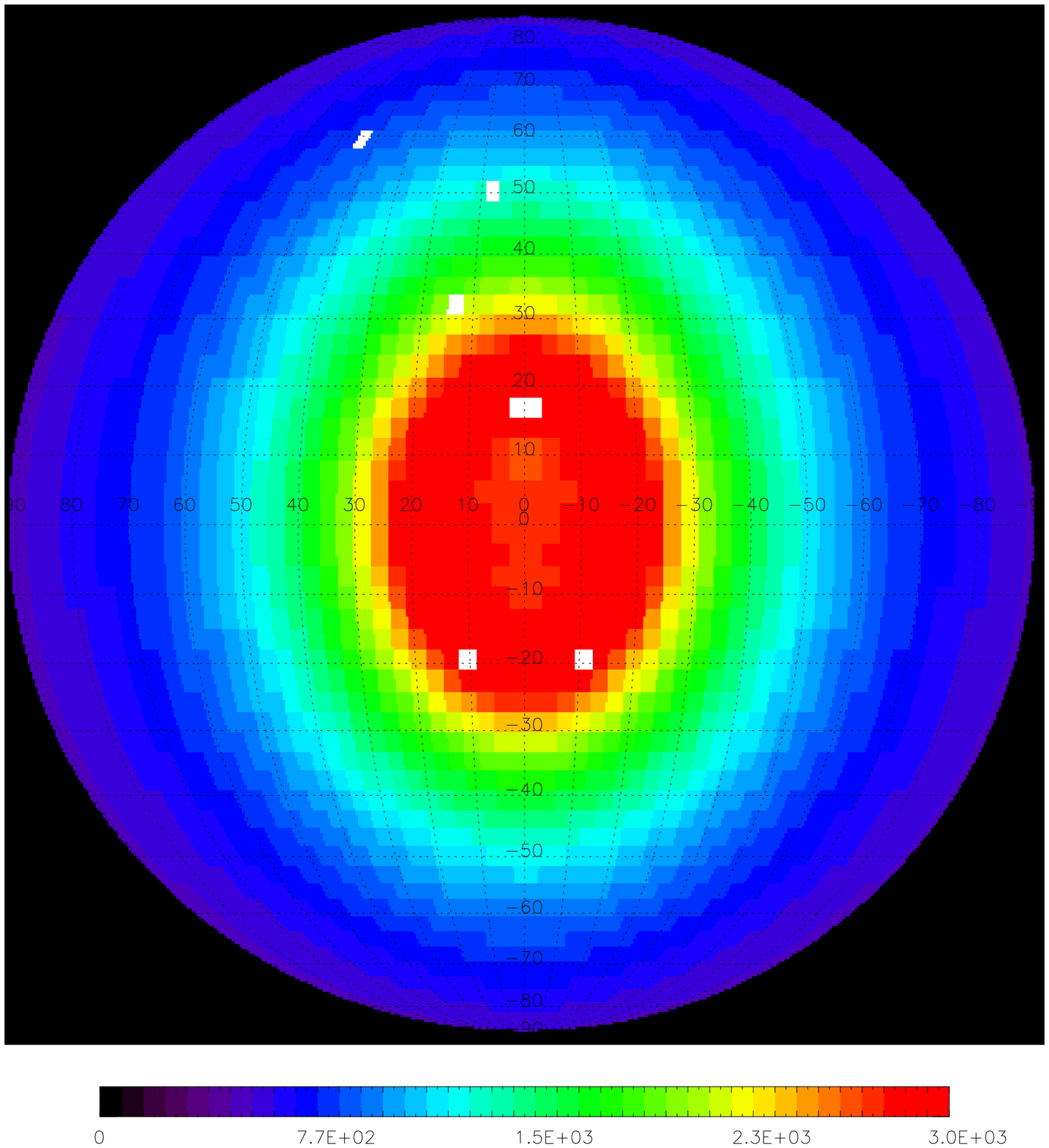}
  {\color{white} \put(-30,20){La,2e}}\\
  \includegraphics[width=0.32\textwidth]{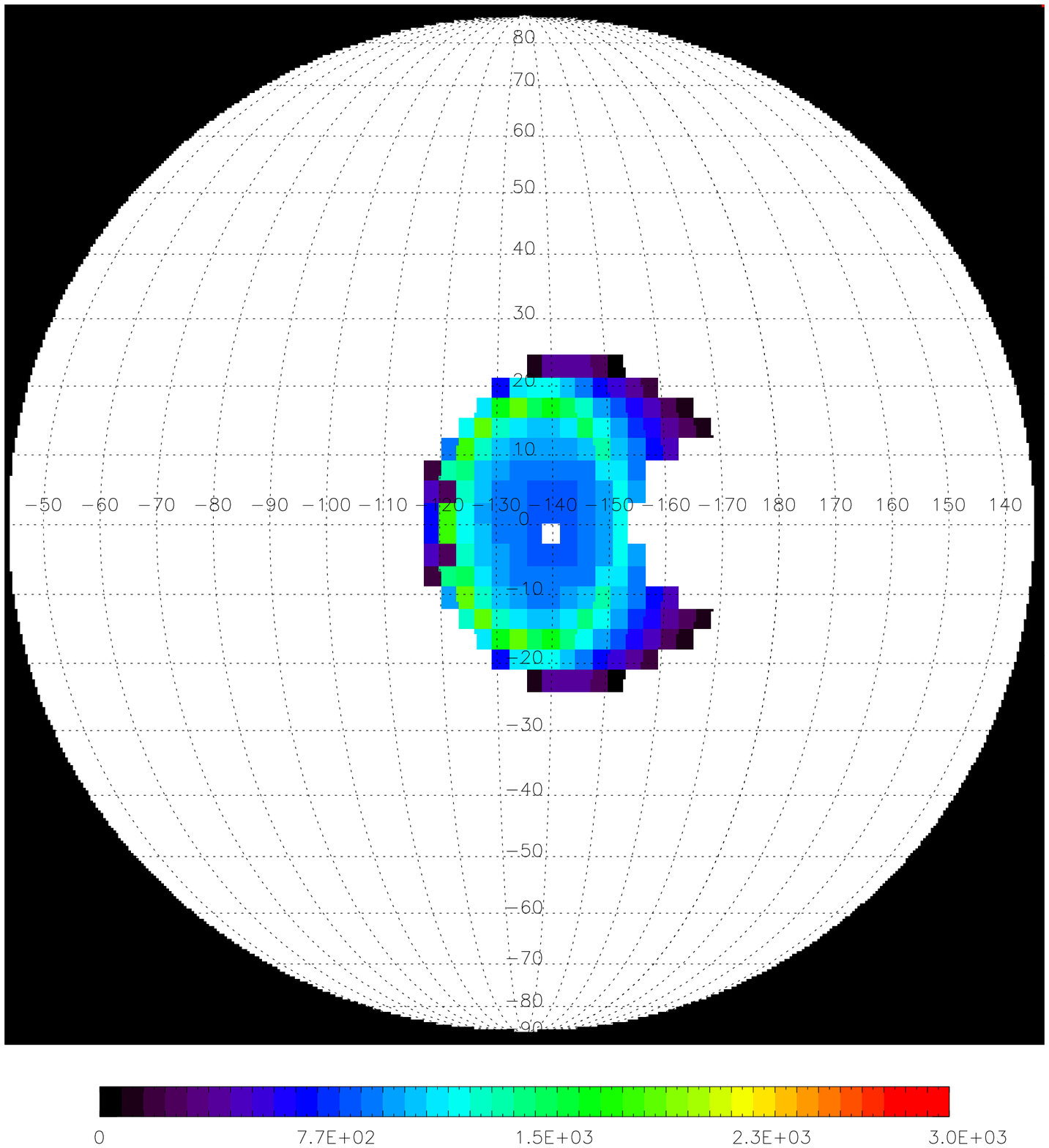}
  {\color{white} \put(-32,20){Cy,2b}}
  \includegraphics[width=0.32\textwidth]{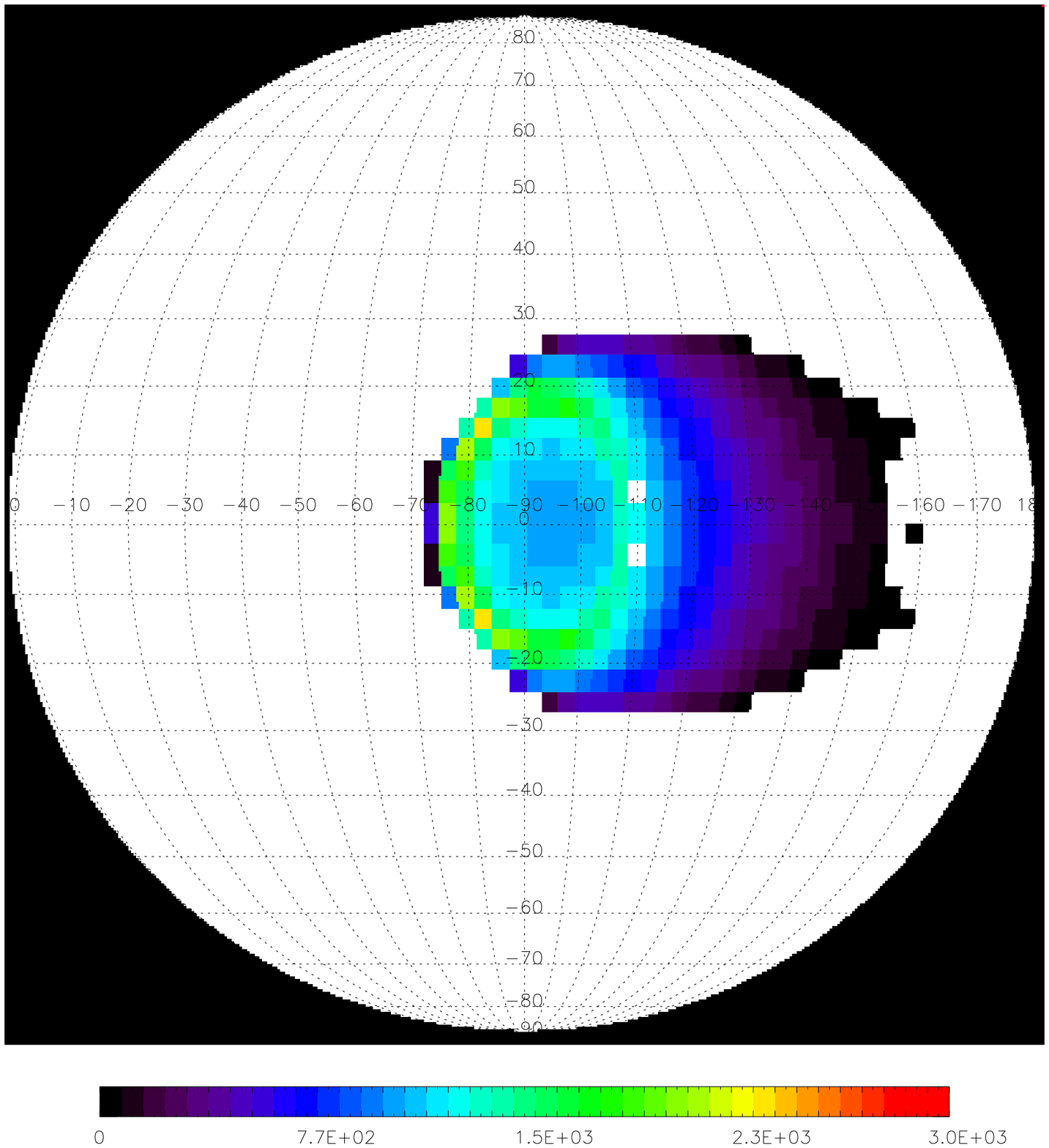}
  {\color{white} \put(-31,20){Cy,2c}}
  \includegraphics[width=0.32\textwidth]{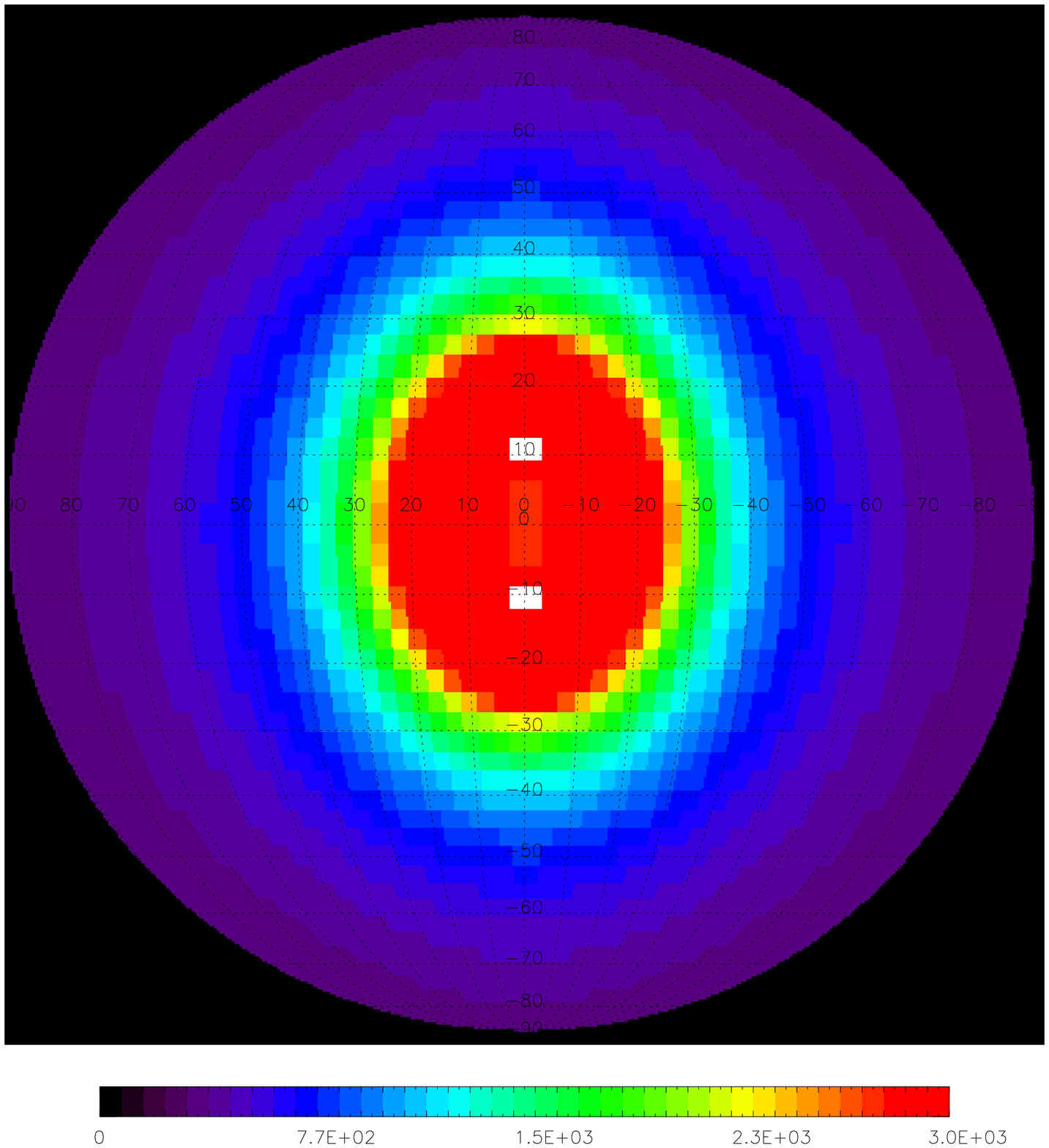}
  {\color{white} \put(-31,20){Cy,2e}}
\caption{ENA map predictions at 400 au heliocentric distance for 
different heliospheric shapes and vantage points for 100 eV ENA energy:
Columns from left to right: intermediate between nose and flank (2b), flank (2c), and downwind observer position (2e);
rows from top to bottom: small ellipsoid, large ellipsoid, and cylindrical shape for heliopause.
The Sun is always in the map center and the color scale is identical across all plots
(red pixels designate ENA intensities $j \ge 3000$ cm$^{-2}$ sr$^{-1}$ s$^{-1}$ keV$^{-1}$).}\label{fig:radial_updown}
\end{figure}

Fig.~\ref{fig:radial_updown} illustrates an advantage for an ENA imager of following a flank trajectory (2c, middle column) 
compared to a rather nose-ward (2b, left column) or tail-ward direction (2e, right column)
as seen from 400 au heliocentric distance and an ENA energy of 100~eV.
Black pixels designate non-detectable ENA intensities ($j<10$ cm$^{-2}$ sr$^{-1}$ s$^{-1}$ keV$^{-1}$ in middle column); the
interstellar ENA background was not modeled and therefore appears white in all plots.  
Different heliospheric shapes, such as the small ellipsoid (top row), large ellipsoid (middle row), and Parker shape (bottom row)
are best distinguished when observed from the flank (middle column).
The mean free path length for 100~eV protons in the heliosheath
is sufficient to make ENAs of the same energy a useful tracer
to image the full extent of the heliopause. The same holds true for ENAs with energies much higher
than 10 keV. However, for a downwind trajectory
(case 2e, $\varphi = 180^{\circ}$, $\vartheta=0^{\circ}$, left column in Fig.~\ref{fig:radial_updown})
even such basic differences in heliosphere shape would be difficult to spot. In addition, 400 au distance towards
downwind direction would not even be sufficient to cross the heliopause 
even if the real heliosphere resembles the small ellipsoid. A trajectory towards the nose obviously does not share
this drawback, but some different heliosphere shapes are impossible to distinguish from the helionose (scenario 2a)
and may still be difficult to distinguish for 2b (cf. top left versus middle left plots in Fig.~\ref{fig:radial_updown}).

From Figs.~\ref{fig:flank_minimum} and \ref{fig:radial_updown}, 
trajectories similar to 2b or 2c appear to offer the most rewarding
vantage points for ENA imaging of the heliosphere.
Trajectory 2b would also offer the opportunity to sample the plasma populations of the Ribbon region in-situ, 
but has less of a novelty value because it is closer to the trajectories of Voyager 1 and 2. Moreover, the heliosphere to the nose direction can be constrained more easily than any other region with ENA observations from the inner solar system (see previous section). Trajectory 2c would offer a better look at the vast downwind regions of the heliosphere in ENA emissions, but reaching the heliopause would take longer and ISN measurements would become more difficult compared to 2b. 

Leaving the ecliptic plane (tested for trajectory 2d with $\vartheta=30^{\circ}$)
changes the apparent position of the Ribbon and the plasma boundaries.
However, these predicted ENA images do not sufficiently differ from the images closer to the ecliptic
at the same longitude to justify the effort of reaching 
higher ecliptic latitudes and the potential loss of science opportunities 
related to Kuiper Belt Objects close to the ecliptic plane.

\begin{figure}
  \includegraphics[width=0.32\textwidth]{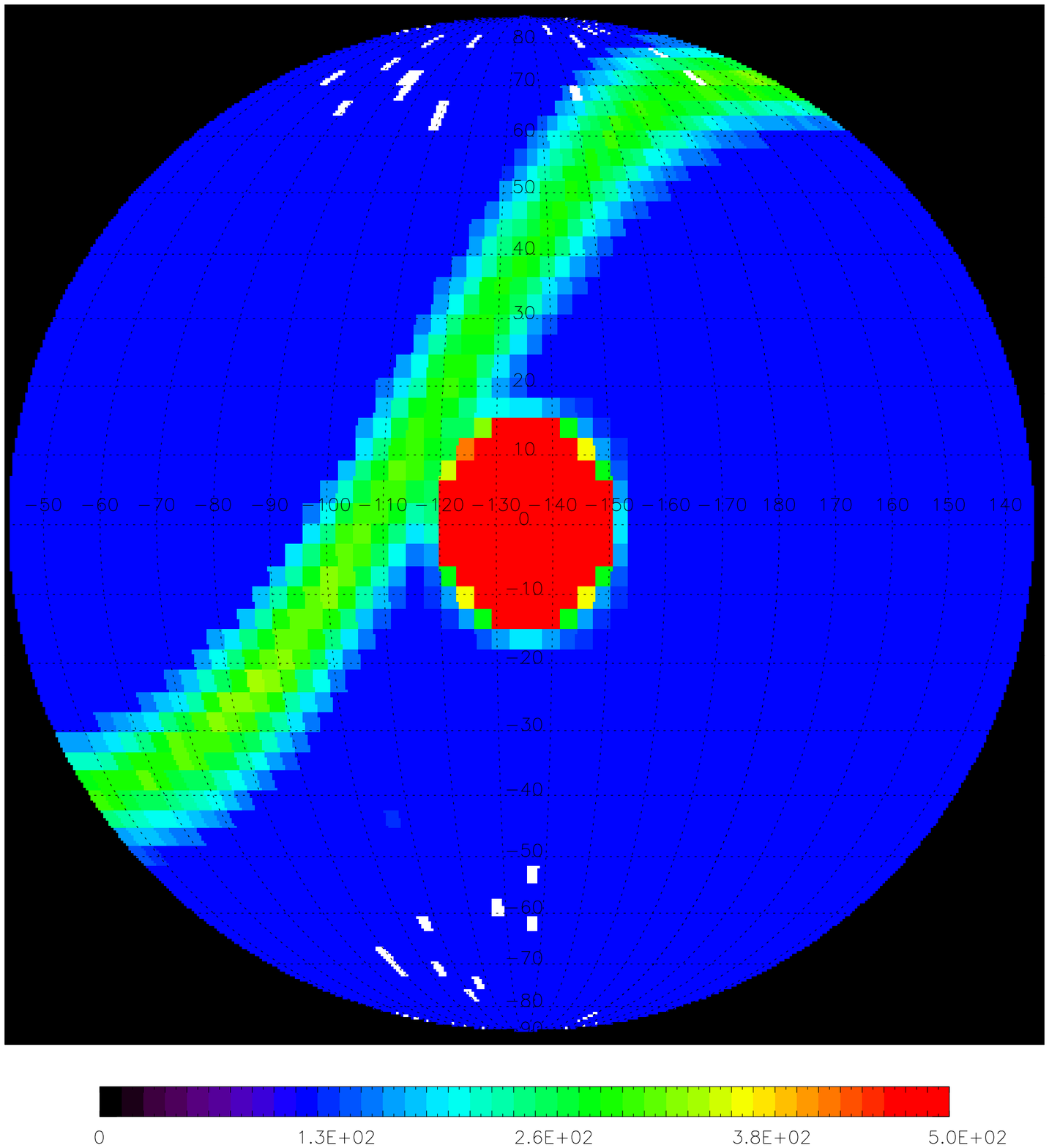}
  {\color{white} \put(-30,20){10 au}}
  \includegraphics[width=0.32\textwidth]{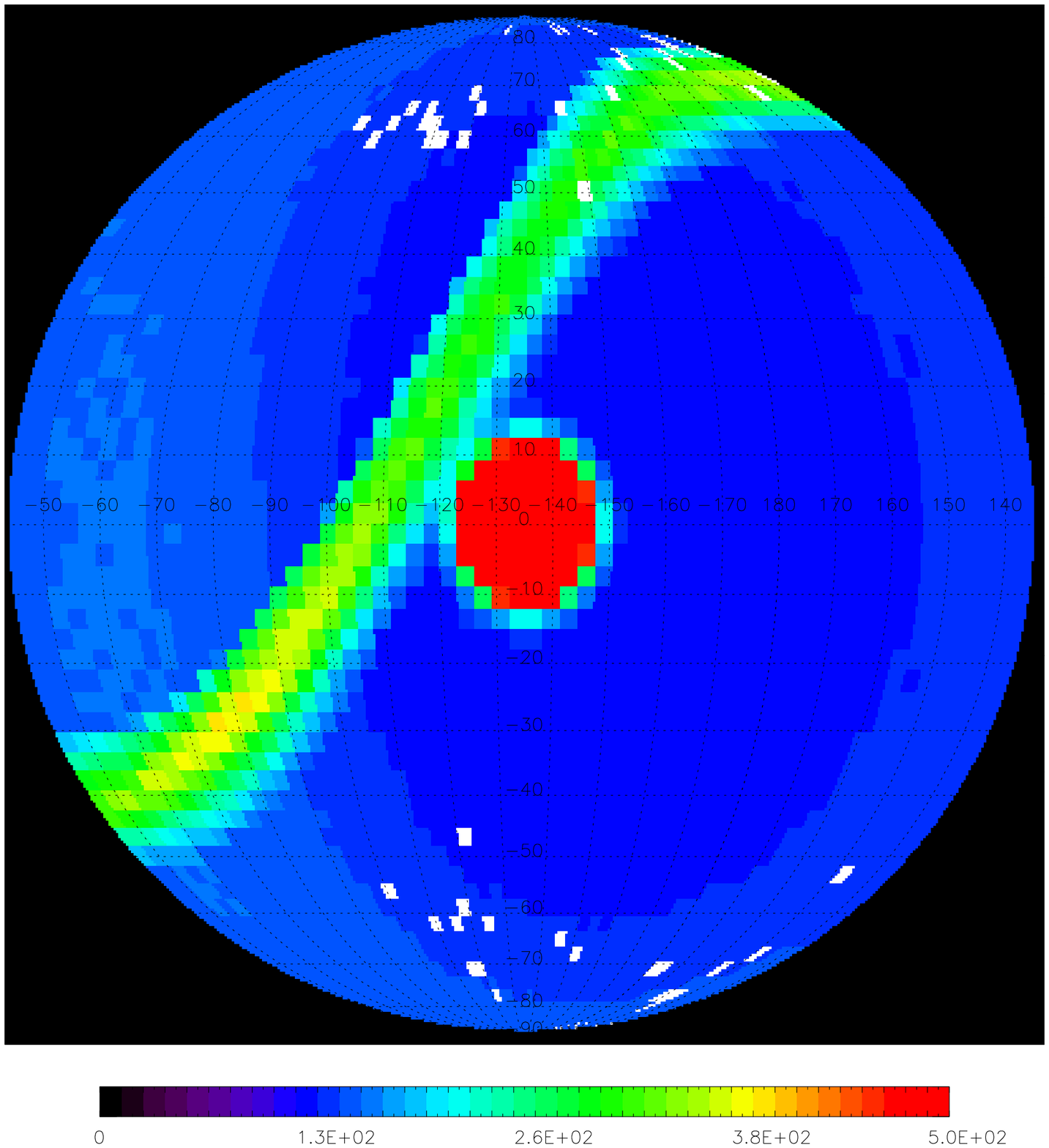}
  {\color{white} \put(-30,20){50 au}}
  \includegraphics[width=0.32\textwidth]{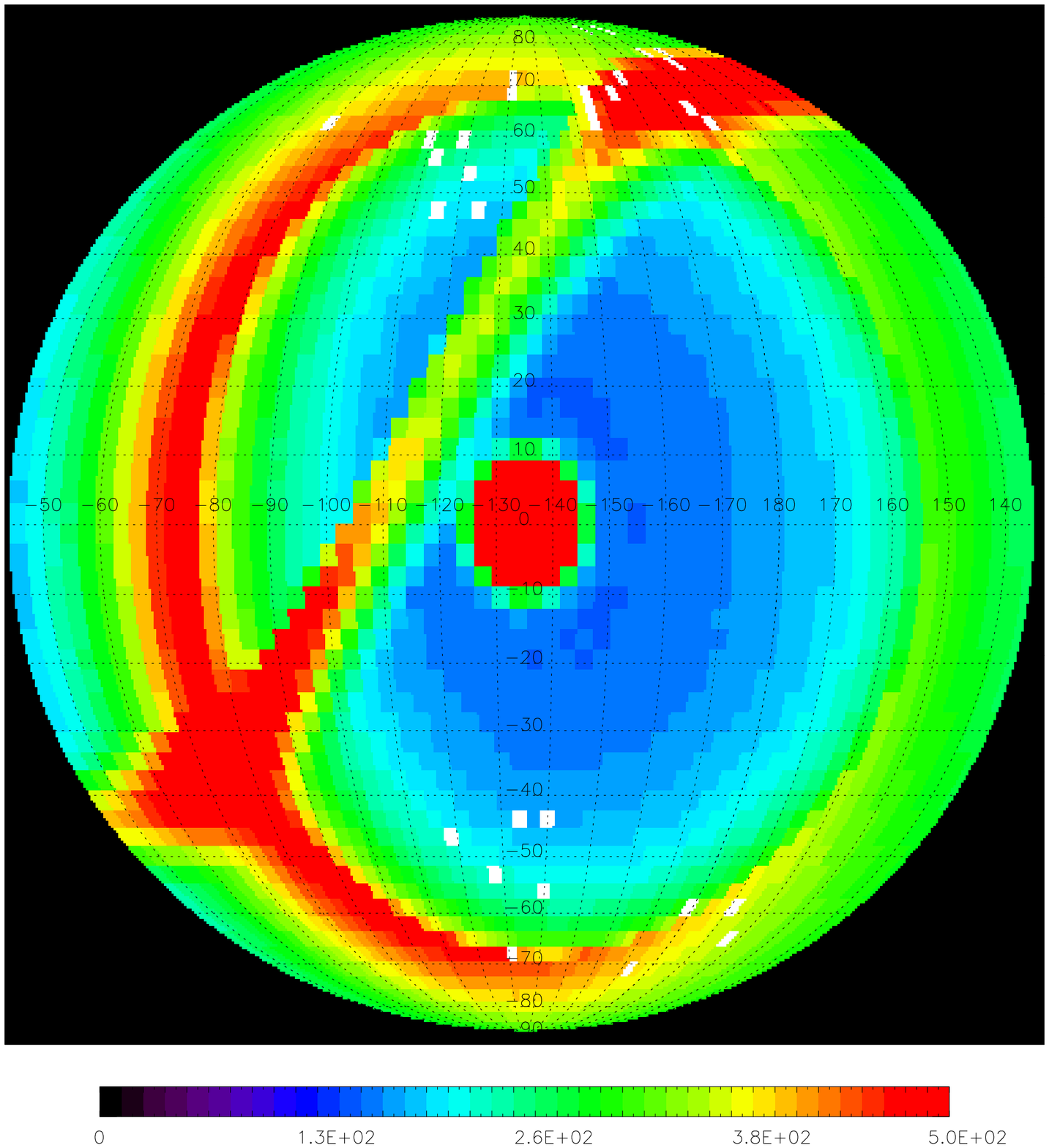}
  {\color{white} \put(-36,20){100 au}}\\
  \includegraphics[width=0.32\textwidth]{2b_422map.eps}
  {\color{white} \put(-36,20){120 au}}
  \includegraphics[width=0.32\textwidth]{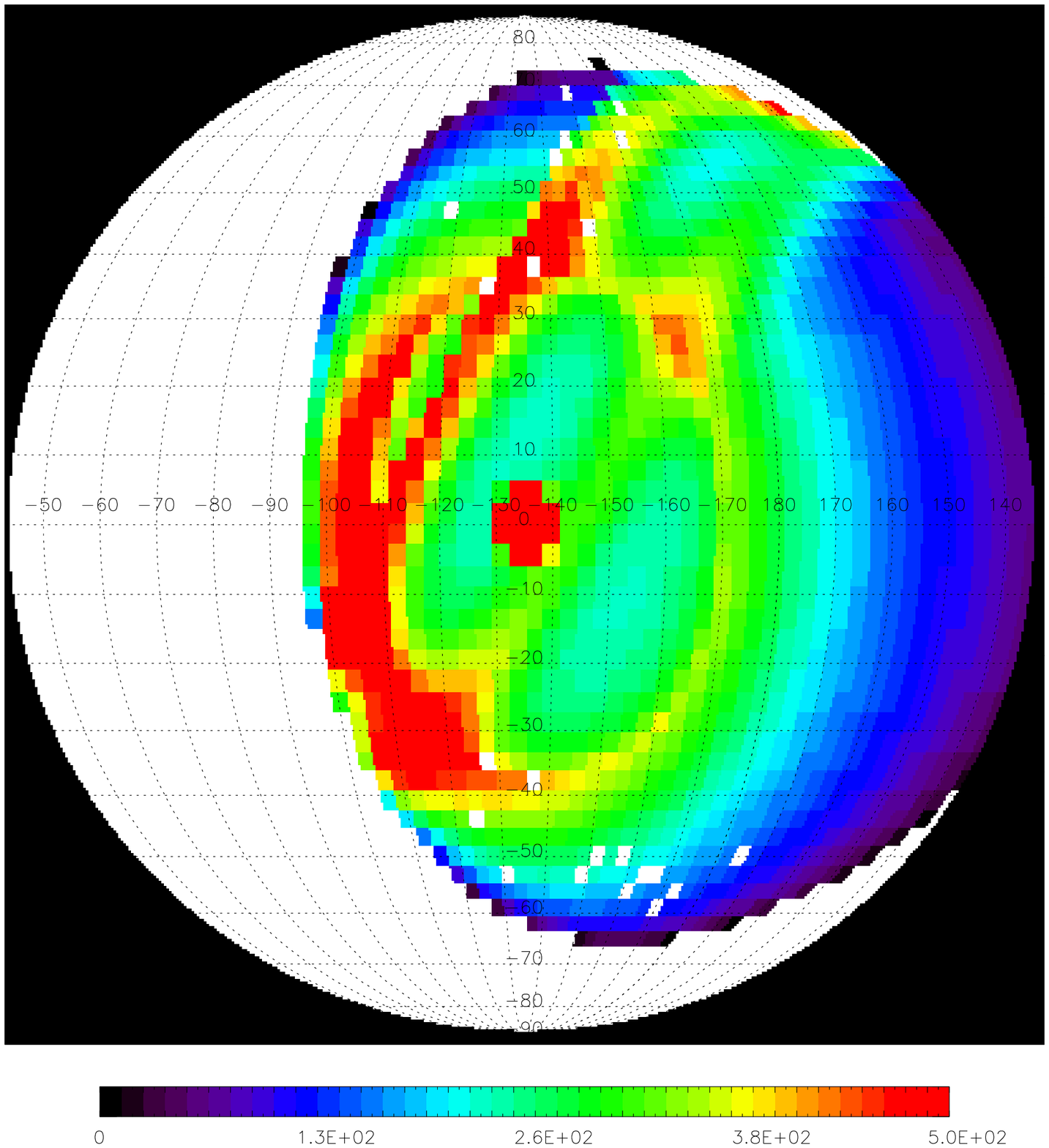}
  {\color{white} \put(-36,20){180 au}}
  \includegraphics[width=0.32\textwidth]{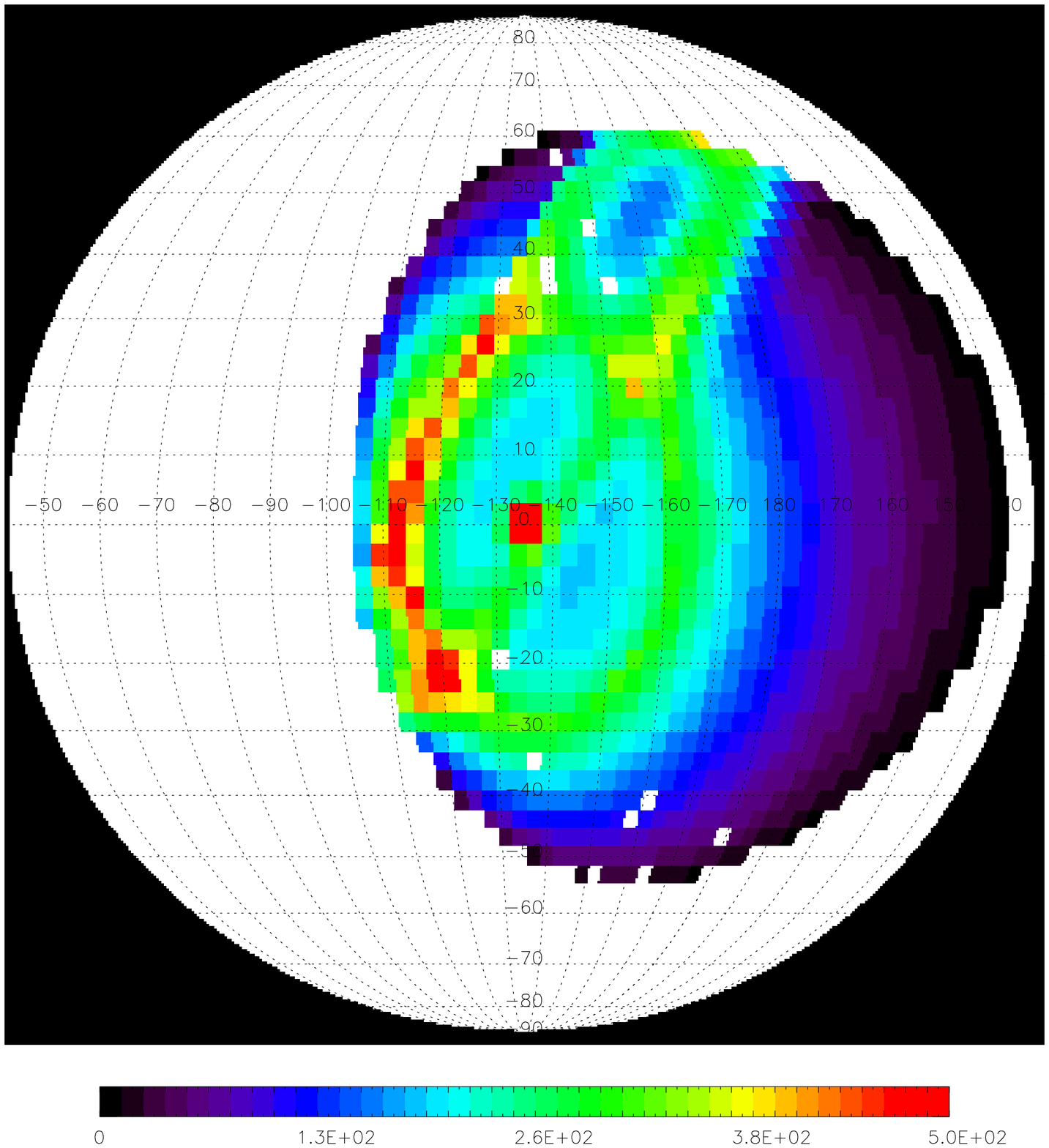}
  {\color{white} \put(-36,20){240 au}}\\
  \includegraphics[width=0.32\textwidth]{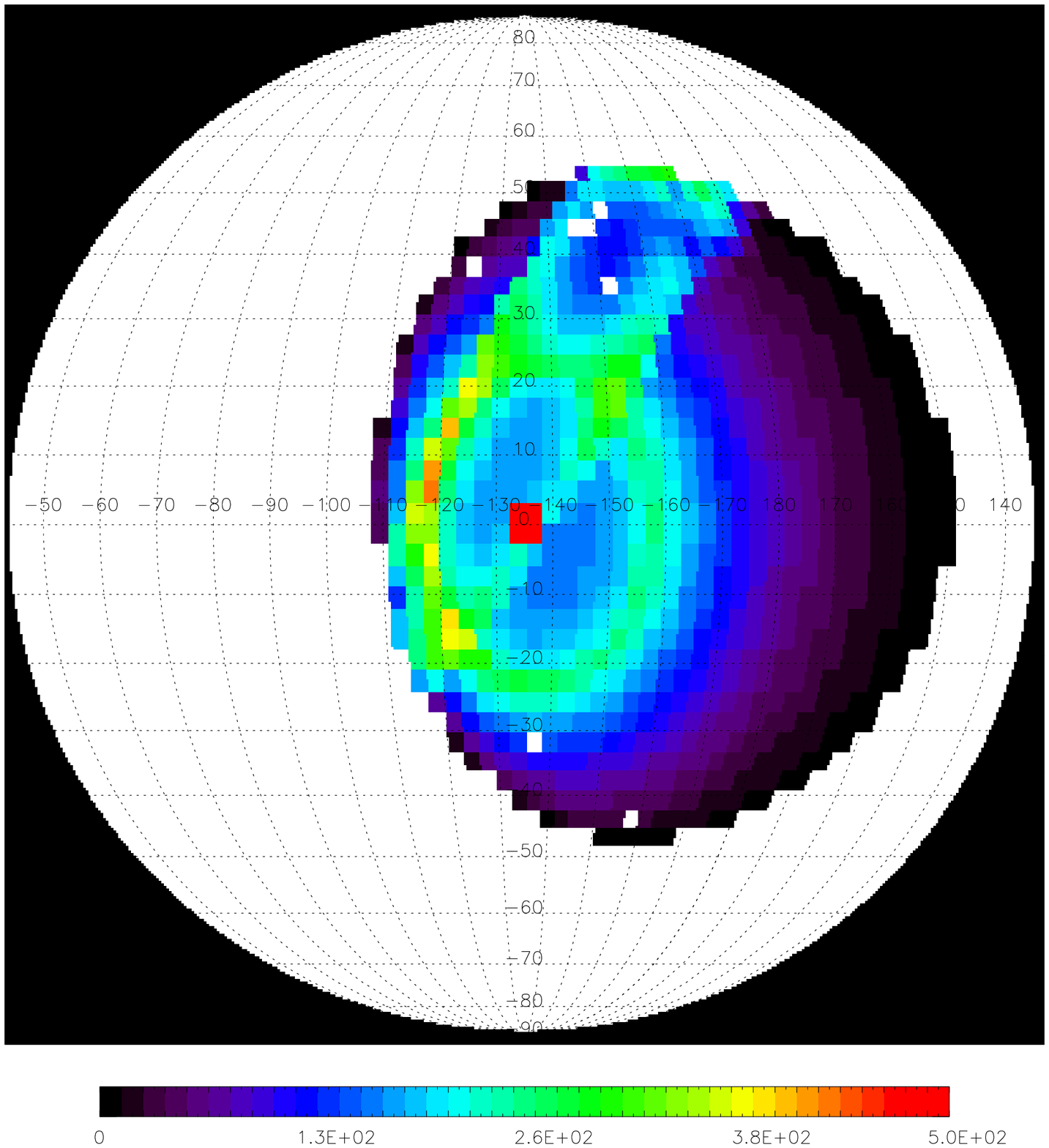}
  {\color{white} \put(-36,20){300 au}}
  \includegraphics[width=0.32\textwidth]{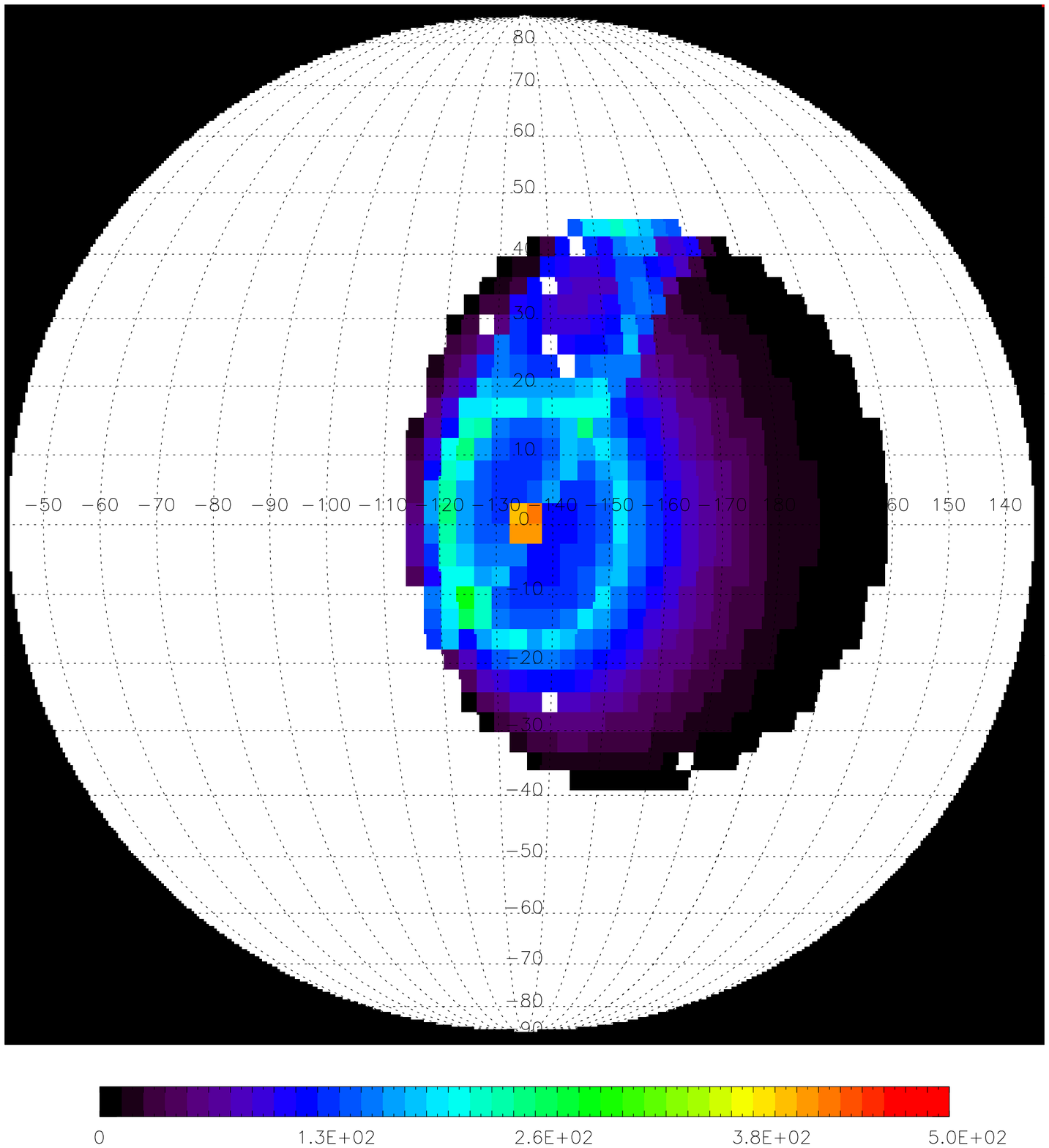}
  {\color{white} \put(-36,20){400 au}}
  \includegraphics[width=0.32\textwidth]{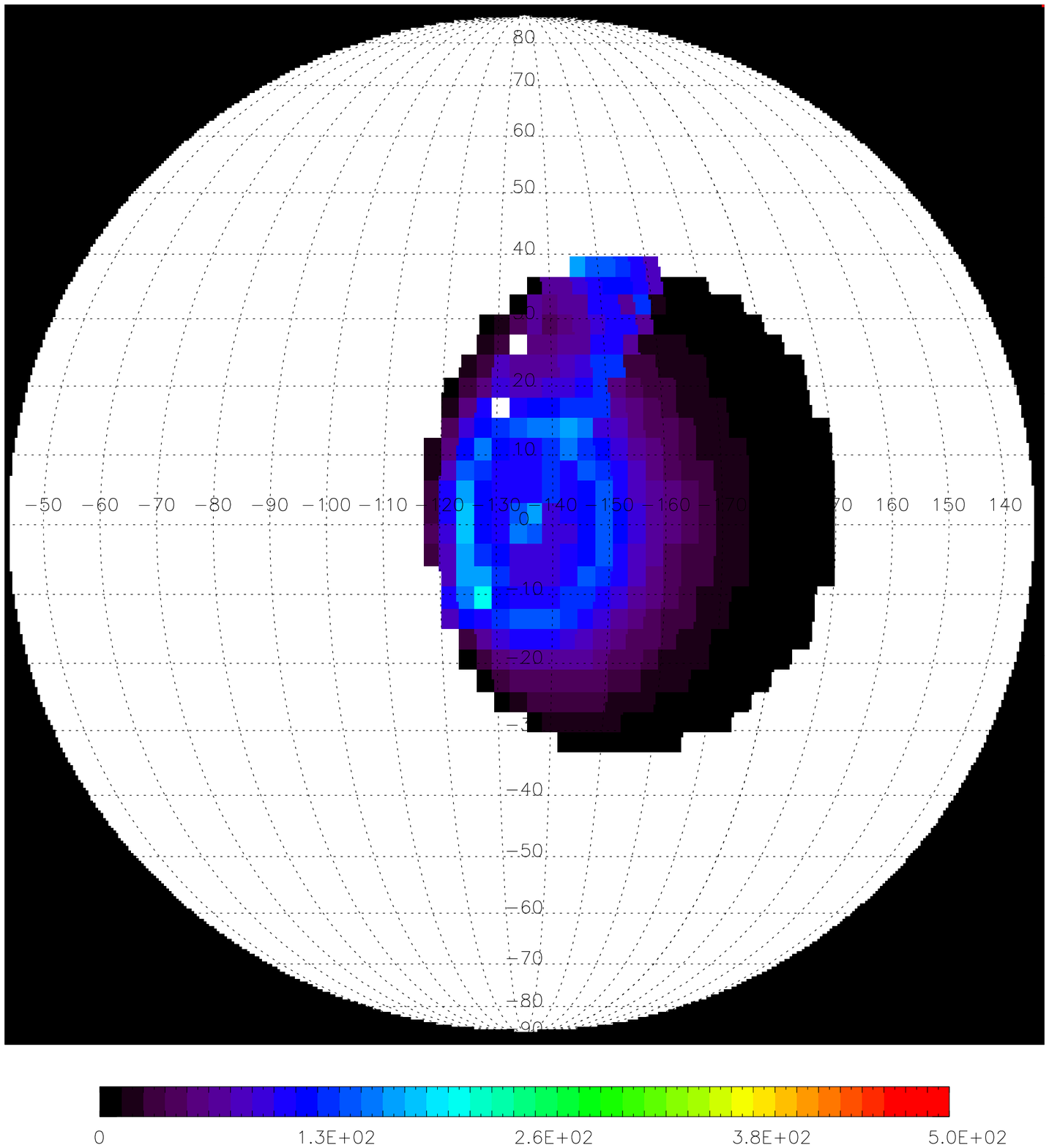}
  {\color{white} \put(-36,20){500 au}}
\caption{Map predictions for 1 keV ENAs for a spacecraft moving radially outwards at a $45^{\circ}$
angle with respect to the nose region (2b), from heliocentric distances 10 to 500 au.
The heliospheric shape assumed here is the large ellipsoid, the color scale for ENA intensity
is identical across all plots (red pixels designate
ENA intensities $j \ge 500$ cm$^{-2}$ sr$^{-1}$ s$^{-1}$ keV$^{-1}$).}\label{fig:radial_flank}
\end{figure}

We finish this section by showing a series of consecutive ENA maps for 1 keV (Fig.~\ref{fig:radial_flank}) 
predicted for a spacecraft leaving the solar system towards 
the flank of the heliosphere (trajectory 2b, $\varphi = 45^{\circ}$, $\vartheta=0^{\circ}$).
The ENA imager is looking back to the Sun while measuring ENAs of 1 keV energy. 
The heliocentric distances increase from top left to bottom right (10, 50, 100, 120, 180, 240, 300, 400, and 500 au), 
the assumed heliosphere shape is the large ellipsoid. Passing through the termination shock 
(roughly at 90 au for this trajectory) dramatically changes the ENA map.
The Ribbon and the neutralized solar wind are prominent features at 1 keV from near and far, 
the GDF from the heliosheath indicates the full dimensions of the heliosphere. Because of the short plasma cooling length
for protons of 1--10 keV in the heliosheath, the farther reaches of the heliosheath in downwind direction are difficult
to image at intermediate energy, but these regions will be visible at lower or higher ENA energies.

\section{Conclusions}\label{sec:conclusions}

We have designed a simple empirical model of heliospheric protons and neutral hydrogen to predict
images of ENA intensities for a virtual observer inside or outside the heliosphere.
The proton distributions are scaled so as to reproduce the known heliospheric
ENA intensities observed close to 1 au. The neutral hydrogen is modeled as a static density, trajectories of interstellar neutrals throughout the heliosphere are not included yet.
We chose this empirical approach to be able to easily visualize 
geometrical effects on heliospheric ENAs for any virtual position and for 
various heliosphere shapes. 

Based on this empirical model and our experience from previous
heliospheric ENA imaging missions, some general recommendation for angular resolution and sensitivity 
of ENA instrumentation have been derived.
All ENA energies from 10 eV to 100 keV have their own merits to characterize
the heliospheric plasma. This wide energy range necessitates two to three different ENA instruments.
Given the uncertainties about the low-energy end of the heliospheric ENA spectrum and 
its relevance for heliosheath properties, a new heliosphere mission should
attempt to image ENAs down to 10 eV. 

For an ENA instrument on a spacecraft orbiting the Sun, the heliocentric distance should be at least
3 au to make use of parallax effects. For an ENA instrument on an interstellar probe, a radial escape trajectory 
through the flank regions of the heliosphere is preferable to central upwind or central downwind direction to image
the global shape of the heliosphere.

\section*{Acknowledgments}

AG and PW acknowledge support from the Swiss National Science Foundation;
HF acknowledges support on ENA-related research via DFG grant FI 706/21-1.

\clearpage

\clearpage

\clearpage

\clearpage

\clearpage

\clearpage

\clearpage

\clearpage

\clearpage

\clearpage

\begin{thebibliography}


\bibitem[Barabash et al.(2006)]{bar06} Barabash, S., et al. 2006, Space Science Reviews, 126, 113

\bibitem[Barabash et al.(2007)]{bar07} Barabash, S., et al. 2007, Planetary and Space Science, 55, 1772


\bibitem[Barabash et al.(2019)]{bar19} Barabash, S., Srama, R., Wieser, M., Wurz, P., \& the LIMO Team 2019, 
``Local Interstellar Medium Observatory (LIMO): a mission to explore our galactic neighborhood'', 
Abstract EGU2019-18299 presented at the EGU conference 2019, Vienna, Austria

\bibitem[Barnett(1990)]{bar90}Barnett, C. F. 1990. Collisions of H, H$_{2}$, He and Li Atoms and Ions
with Atoms and Molecules, Atomic Data for Fusion, Vol. 1, Oak Ridge National Laboratory,
Report ORNL-6086 


\bibitem[Brandt et al.(2019)]{bra19} Brandt, P. C., McNutt, R. L., Paul, M. V., et al. 2019, 
``Humanity's First Explicit Step In Reaching Another
Star: The Interstellar Probe Mission'', Journal of the British Interplanetary Society, \textit{submitted}

\bibitem[Burlaga et al.(2005)]{bur05} Burlaga, L. F., Ness, N. F., Acu\~{n}a, M. H., et al.
2005, Science 309, 2027

\bibitem[Burlaga et al.(2008)]{bur08} Burlaga, L. F., Ness, N. F., Acu\~{n}a, M. H., et al.
2008, Nature, 454, 75

\bibitem[Bzowski(2008)]{bzo08} Bzowski, M. 2008, Astronomy \& Astrophysics, 488, 1057




\bibitem[Bzowski et al.(2013)]{bzo13} Bzowski, M., Sok\'{o}{\l}, J. M., Kubiak, M. A., \& Kucharek, H. 2013, Astronomy \& Astrophysics, 557, A50




\bibitem[Dayeh et al.(2019)]{day19} Dayeh, M. A., Zirnstein, E. J., Desai, M. I., et al. 2019, The Astrophysical Journal, 879, 84


\bibitem[DeMajistre et al.(2018)]{dem18} DeMajistre, R., Brandt P. C., Roelof, E. C. 2018, 
Looking back at our heliosphere in ENA,	42nd COSPAR Scientific Assembly, abstract id. PIR.1-27-18



\bibitem[Desai et al.(2019)]{des19} Desai, M. I., Dayeh, M. A., Allegrini, F., et al. 2019, The Astrophysical Journal, 875, 91

\bibitem[Dialynas et al.(2017)]{dia17} Dialynas, K., Krimigis, S. M., Mitchell, D. G., Decker, R. B., \& Roelof, E. C. 2017,
Nature Astronomy, 1, 115


\bibitem[Elliott et al.(2016)]{ell16} Elliott, H. A., McComas, D. J., Valek, P., Nicolaou, G., Weidner, S., \& Livadiotis, G. 2016,
The Astrophysical Journal Supplement Series, 223, 19

\bibitem[Fahr et al.(1986)]{fah86} Fahr, H. J., Neutsch, W., Grzedzielski, S., Macek, W., \& Ratkiewicz-Landowska, R. 1986,
Space Science Reviews, 43, 329

\bibitem[Fahr et al.(2007)]{fah07} Fahr, H. J., Fichtner, H., \& Scherer, K. 2007,
Reviews of Geophysics, 45, RG4003



\bibitem[Funsten et al.(2009)]{fun09} Funsten, H. O., Allegrini, F., Bochsler, P., 
et al. 2009, Space Science Reviews, 146, 75

\bibitem[Fuselier \& Pope(2005)]{fus05}Fuselier, S.A., \& Pope, S. 2005, ``Interstellar Boundary EXplorer (IBEX) noise and back-
ground document'', SwRI Project 11343, Document No. 11343-IBEX BKGND-01

\bibitem[Fuselier et al.(2009)]{fus09} Fuselier, S. A., Bochsler, P., Chornay, D.,
et al. 2009, Space Science Reviews, 146, 117


\bibitem[Fuselier et al.(2012)]{fus12} Fuselier, S. A., Allegrini, F., Bzowski, M., et al. 2012, The Astrophysical Journal, 754, 14

\bibitem[Fuselier et al.(2014)]{fus14} Fuselier, S. A., Allegrini, F.,
Bzowski, M., et al. 2014, The Astrophysical Journal, 784, 89

\bibitem[Fuselier et al.(2018)]{fus18} Fuselier, S. A., Dayeh, M. A., \& Möbius, E. 2018, The Astrophysical Journal, 861, 109

\bibitem[Futaana et al.(2006)]{fut06} Futaana, Y., Barabash, S., Grigoriev A., et al. 2006, Icarus, 182, 413


\bibitem[Galli et al.(2013)]{gal13} Galli, A., Wurz, P., Kollmann, P., et al.
2013, The Astrophysical Journal, 775, 24

\bibitem[Galli et al.(2014)]{gal14} Galli, A., Wurz, P., Fuselier, S. A., et al.
2014, The Astrophysical Journal, 796, 9


\bibitem[Galli et al.(2016)]{gal16} Galli, A., Wurz, P., Schwadron, N.A., et al.
2016, The Astrophysical Journal, 821, 107

\bibitem[Galli et al.(2017)]{gal17} Galli, A., Wurz, P., Schwadron, N.A., et al.
2017, The Astrophysical Journal, 851, 2

\bibitem[Galli et al.(2019)]{gal19} Galli, A., Wurz, P., Rahmanifard, F., et al.
2019, The Astrophysical Journal, 871, 52

\bibitem[Giacalone and Jokipii(2015)]{gia15} Giacalone, J., \& Jokipii, J. R. 2015, The Astrophysical Journal Letters, 812, 1


\bibitem[Gloeckler et al.(2004)]{glo04}Gloeckler, G., Allegrini, F., Elliott, H. A., et al. 2004, The Astrophysical Journal Letters, 604, L121

\bibitem[Gloeckler \& Fisk(2011)]{glo11} Gloeckler, G. \& Fisk, L. A. 2011, in AIP Conf. Proc. 1302, Pickup Ions
throughout the Heliosphere and Beyond, ed. J. A. Le Roux, V. Florinski, G.
P. Zank, \& A. J. Coates (Melville, NY: AIP), 110, doi: http://dx.doi.org/10.1063/1.3529957

\bibitem[Gloeckler \& Fisk(2015)]{glo15} Gloeckler, G. \& Fisk, L. A. 2015, The Astrophysical Journal Letters, 806, L27

\bibitem[Gosling(2007)]{gos07}Gosling, J.T., 2007, ``Chapter 5 -- The Solar Wind'', in: Encyclopedia of the Solar System (Second Edition), 
eds. L.A. McFadden, P.R. Weissman, \& T.V. Johnson, (San Diego, CA: Academic Press), 99--116, 
doi.org/10.1016/B978-012088589-3/50009-8


\bibitem[Gruntman et al.(2001)]{gru01} Gruntman, M., Roelof, E. C., Mitchell, et al.
2001, \jgr, 106, 15767

\bibitem[Heerikhuisen et al.(2014)]{hee14} Heerikhuisen, J., Zirnstein, E. J., Funsten, H. O., Pogorelov, N. V., \& Zank, G.
P. 2014, The Astrophysical Journal, 784, 73

\bibitem[Hilchenbach et al.(1998)]{hil98} Hilchenbach, M., Hsieh, K. C., Hovestadt, D., et al.
1998, The Astrophysical Journal, 503, 916

\bibitem[Ipavich(1974)]{ipa74} Ipavich, F. M. 1974, Geophysical Research Letters, 1, 149


\bibitem[Izmodenov \& Alexashov(2015)]{izm15} Izmodenov, V. V. \& Alexashov, D. B. 2015, 
The Astrophysical Journal Supplement Series, 220, 32

\bibitem[Kallenbach et al.(2005)]{kal05} Kallenbach, R., Hilchenbach, M., Chalov, S. V., 
Le Roux, J. A., \& Bamert, K. 2005, Astronomy \& Astrophysics, 439, 1 


\bibitem[Khabarova et al.(2018)]{kha18} Khabarova, O. V. Obridko, V. N., Kislov, R. A., Malova, H. V., Bemporad, A.,
Zelenyi, L. M., Kuznetsov, V. D., \& Kharshiladze, A. F. 2018, Plasma Physics Reports, 44, 840




\bibitem[Krimigis et al.(2009)]{kri09}Krimigis, S. M., Mitchell, D. G., Roelof, E. C., Hsieh, K. C., \& McComas, D. J. 2009, Sci, 326, 971


\bibitem[Kubiak et al.(2014)]{kub14} Kubiak, M. A., Bzowski, M., Sok\'{o}{\l}, J. M., et al. 2014, The Astrophysical Journal Supplement Series,
213, 29



\bibitem[Lindsay \& Stebbings(2005)]{lin05} Lindsay, B. G. \& Stebbings, R. F. 2005, Journal of Geophysical Research, 110, A12213

\bibitem[Livadiotis et al.(2013)]{liv13} Livadiotis, G., McComas, D. J., Schwadron, N. A., Funsten, H. O., \& Fuselier,
S. A. 2013, The Astrophysical Journal, 762, 134

\bibitem[Marsch et al.(1982)]{mar82} Marsch, E., M\"uhlh\"auser, K.-H., Schwenn, R., et al.
1982, \jgr, 87, 52




\bibitem[McComas et al.(2010)]{mcc10} McComas, D. J., Bzowski, M., Frisch, P., et al.
2010, \jgr, 115, A09113, doi:10.1029/2010JA015569

\bibitem[McComas et al.(2012)]{mcc12}McComas, D. J., Alexashov, D., Bzowski, M., et al. 2012, Science, 336, 1291




\bibitem[McComas et al.(2014)]{mcc14} McComas, D. J., Lewis, W. S., \& Schwadron, N. A. 2014,
Reviews of Geophysics, 52, doi:10.1002/2013RG000438

\bibitem[McComas et al.(2015)]{mcc15} McComas, D. J., Bzowski, M., Fuselier, S. A., et al. 2015, The Astrophysical Journal Supplement Series, 220, 22

\bibitem[McComas et al.(2017)]{mcc17} McComas, D. J., Zirnstein, E. J., Bzowski, M., et al. 2017, The Astrophysical Journal Supplement Series, 229, 41

\bibitem[McComas et al.(2018)]{mcc18} McComas, D. J., Christian, E. R., Schwadron, N. A., et al. 2018, Space Science Reviews, 214, 116

\bibitem[McComas et al.(2019)]{mcc19} McComas, D. J., Dayeh, M., A., Funsten, H. O., et al. 2019, The Astrophysical Journal, 872, 127

\bibitem[McNutt et al.(2018)]{mcn18} McNutt, R. L., Wimmer-Schweingruber, R. F., Gruntman, M., et al. 2018,
``Near-Term Interstellar Probe: First Step'', presented at the 69th International Astronautical Congress, Bremen, Germany


\bibitem[M\"{o}bius et al.(2012)]{moe12} M\"{o}bius, E., Bochsler, P., Bzowski, M., et al. 2012,
The Astrophysical Journal Supplement Series, 198, 11


\bibitem[M\"uller \& Zank(2004)]{mue04}M\"uller, H.-R. \& Zank, G. P. 2004, Journal of Geophysical Research 109, A07104

\bibitem[Opher(2016)]{oph16a} Opher, M. 2016, Space Science Reviews, 200, 475

\bibitem[Opher et al.(2016)]{oph16} Opher, M., Drake, J. F., Zieger, B., Swisdak, M., \& Toth, G. 2016,
Physics of Plasmas 23, 056501

\bibitem[Park et al.(2016)]{par16} Park, J., Kucharek, H., M\"{o}bius, E., et al. 2016, The Astrophysical Journal, 833, 130

\bibitem[Parker(1961)]{par61} Parker, E. N., 1961, The Astrophysical Journal, 134, 20 


\bibitem[Pogorelov et al.(2017)]{pog17} Pogorelov, N. V., Fichtner, H., Czechowski, A., et al. 2017, 
Space Science Reviews, 212, 193-248

\bibitem[Rahmanifard et al.(2019)]{rah19} Rahmanifard, F., M\"obius, E., Schwadron, N. A., et al. 2019, 
\textit{submitted to the Astrophysical Journal}


\bibitem[Reisenfeld et al.(2016)]{rei16} Reisenfeld, D. B., Bzowski, M., Funsten, H.O. et al.
2016, The Astrophysical Journal, 833, 277

\bibitem[Reisenfeld et al.(2019)]{rei19} Reisenfeld, D. B., Bzowski, M., Funsten, H.O. et al.
2019, The Astrophysical Journal, 879, 1


\bibitem[Rodr\'{i}guez et al.(2012)]{rod12} Rodr\'{i}guez Moreno, D. F., Saul, L., Wurz, P., et al.
2012, Planetary and Space Science, 60 297 

\bibitem[R\"{o}ken et al.(2015)]{roe15} 
R\"{o}ken, C., Kleimann, J., \& Fichtner, H., 2015, The Astrophysical Journal, 805, 173, 2015

\bibitem[Roelof et al.(1976)]{roe76} Roelof, E. C., Keath, E. P., \& Bostrom, C. O. 1976, 
Journal of Geophysical Research, 81, 2304

\bibitem[Roelof et al.(1985)]{roe85} Roelof, E. C., Mitchell, D. G., \& Williams, D. J., 1985, Journal of Geophysical Research, 90, 10991


\bibitem[Saul et al.(2012)]{sau12} Saul, L., Wurz, P., Rodr\'{i}guez, D., et al. 2012, The Astrophysical Journal Supplement Series, 198, 14


\bibitem[Scherer and Fichtner(2014)]{sche14} Scherer, K., \& Fichtner, H. 2014, The Astrophysical Journal, 782, 25

\bibitem[Schwadron et al.(2011)]{sch11} Schwadron, N. A., Allegrini, F., Bzowski, M., et al.
2011, The Astrophysical Journal, 731, 56


\bibitem[Schwadron et al.(2014)]{sch14} Schwadron, N. A., M\"{o}bius, E., Fuselier, S. A. et al.
2014, The Astrophysical Journal, 215, 13


\bibitem[Schwadron and Bzowski(2018)]{sch18} Schwadron, N. A., \& Bzowski, M. 2018, The Astrophysical Journal, 862, 11






\bibitem[Stone et al.(2013)]{sto13} Stone, E. C., Cummings, A. C., McDonald, F. B., et al. 2013, Science, 341, 15



\bibitem[Swaczyna et al.(2016)]{swa16} Swaczyna, P., Bzowski, M., Christian, E. R., et al. 2016, The Astrophysical Journal, 823, 119

\bibitem[Swaczyna et al.(2017)]{swa17} Swaczyna, P., Grzedzielski, S., \& Bzowski, M. 2017, The Astrophysical Journal, 840, 75


\bibitem[Sylla and Fichtner(2015)]{syl15} Sylla, A., Fichtner, H. 2015, The Astrophysical Journal, 811, 150



\bibitem[Witte(2004)]{wit04}Witte, M., 2004, Astronomy\&Astrophysics, 426, 835 




\bibitem[Wurz(2000)]{wur00} Wurz, P., 2000, Detection of energetic neutral particles, in 
``The Outer Heliosphere: Beyond the Planets'', eds. K. Scherer, H. Fichtner, E. Marsch 
(Copernicus Gesellschaft e.V., Katlenburg-Lindau, 2000), 251


\bibitem[Wurz et al.(2009)]{wur09}Wurz, P., et al., 2009. Space Science Reviews, 146, 173




\bibitem[Zank et al.(2013)]{zan13} Zank, G. P., Heerikhuisen, J., Wood, B. E., Pogorelov, N. V., Zirnstein, E.,
 \& McComas, D. J. 2013, The Astrophysical Journal, 763, 20

\bibitem[Zirnstein et al.(2016a)]{zir16a} Zirnstein, E. J., Funsten, H. O., Heerikhuisen, J., \& McComas, D. J. 2016a, 
Astronomy \& Astrophysics 586, A31

\bibitem[Zirnstein et al.(2016b)]{zir16b} Zirnstein, E. J., Funsten, H. O., Heerikhuisen, J., McComas, D. J., Schwadron, N. A., \& Zank, G. P. 2016b, The Astrophysical Journal, 826, 58

\bibitem[Zirnstein et al.(2016c)]{zir16c} Zirnstein, E. J., Heerikhuisen, J., Funsten, H. O., Livadiotis, G., McComas, D. J., \&
Pogorelov, N. V. 2016c, The Astrophysical Journal Letters, 818, L18


\bibitem[Zirnstein et al.(2018)]{zir18} Zirnstein, E. J., Kumar, R., Heerikhuisen, J., McComas, D. J., \& Galli, A. 2018, The Astrophysical Journal, 860, 170

\bibitem[Zirnstein et al.(2019)]{zir19} Zirnstein, E. J., Giacalone, J., et al. 2019, The Astrophysical Journal, \textit{submitted}

\bibitem[Zong(2018)]{zon18} Zong, Q.-G., 2018, ``Interstellar Heliosphere Probes'', 
Abstract PIR.1-0002-18 presented at the 42th COSPAR assembly, Pasadena, USA

\end{thebibliography}
\end{document}